\documentclass[aps,prb,10pt,twocolumn,superscriptaddress,floatfix,nofootinbib,longbibliography,amsmath,amssymb]{revtex4-2}

\usepackage{amsfonts}
\usepackage{braket}

\usepackage[dvipsnames]{xcolor}
\usepackage{graphicx}
\usepackage{tikz}

\usepackage{float}
\usepackage{subfigure}
\usepackage{sidecap}

\usepackage[normalem]{ulem}
\usepackage{xspace}

\usepackage[colorlinks=true,linktoc=page,linkcolor=Purple,citecolor=RedViolet,urlcolor=Magenta]{hyperref}

\newif\ifdraft
\drafttrue
\ifdraft
  
\else

\fi
\newcommand{\rc}{r_c}   % core radius of the out-of-plane spin component (was $a$); black
\newcommand{\Nsk}{N_{\mathrm{sk}}}   % topological charge / skyrmion number (was $\mathcal{N}$); black

\newcommand{\eg}{e.g.,\xspace}

\mathchardef\mhyphen="2D           % math hyphen
\newcommand{\bea}{\begin{eqnarray}}
\newcommand{\eea}{\end{eqnarray}}
\newcommand{\beq}{\begin{equation}}
\newcommand{\eeq}{\end{equation}}
\definecolor{lime}{HTML}{A6CE39}
\DeclareRobustCommand{\orcidicon}{%
    \hspace{0.2mm}%
    \begin{tikzpicture}
	\draw[lime, fill=lime] (0.0,0.0) circle [radius=0.15] 
	   node[white] {{\fontfamily{qag}\selectfont \tiny \,ID}};
	\draw[white, fill=white] (-0.0525,0.095) 
	   circle [radius=0.007];
    \end{tikzpicture}%
    \hspace{0.2mm}}
    
\foreach \x in {A,...,Z}{%
  \expandafter\xdef\csname orcid\x\endcsname{%
    \noexpand\href{https://orcid.org/\csname orcidauthor\x\endcsname}%
    {\noexpand\orcidicon}}}

\makeatletter

\AtBeginDocument{%
    \newwrite\bibnotes
    \def\bibnotesext{Notes.bib}%
    \immediate\openout\bibnotes=\jobname\bibnotesext
    \immediate\write\bibnotes{@CONTROL{REVTEX42Control}}
    \immediate\write\bibnotes{@CONTROL{apsrev41Control,%
     author="08",editor="1",pages="1",title="1",year="1"}}
    \if@filesw
    \immediate\write\@auxout{\string\citation{apsrev42Control}}%
    \fi
}

\makeatother

\begin{document}

	%=============START of MAIN PAPER===============

%=================== Title===========================================
\title{Tunable Majorana corner modes in a topological insulator$-$superconductor heterostructure overlaid with magnetic meron textures} 
%====================================================================
	
	\author{Ohidul Alam\orcidA{}}
	\email{ohidul.alam@iopb.res.in}
	\affiliation{Institute of Physics, Sachivalaya Marg, Bhubaneswar-751005, India}
	\affiliation{Homi Bhabha National Institute, Training School Complex, Anushakti Nagar, Mumbai 400094, India}
	
	\author{Bhairab Mondal\orcidB{}}
	\affiliation{School of Physical Sciences, National Institute of Science Education and Research, Jatni 752050, India}
	\affiliation{Homi Bhabha National Institute, Training School Complex, Anushakti Nagar, Mumbai 400094, India}

	\author{Arijit Saha\orcidD{}}
	\affiliation{Institute of Physics, Sachivalaya Marg, Bhubaneswar-751005, India}
	\affiliation{Homi Bhabha National Institute, Training School Complex, Anushakti Nagar, Mumbai 400094, India}
	
	\author{Ashis K. Nandy\orcidC{}}
	\email{aknandy@niser.ac.in}
	\affiliation{School of Physical Sciences, National Institute of Science Education and Research, Jatni 752050, India}
	\affiliation{Homi Bhabha National Institute, Training School Complex, Anushakti Nagar, Mumbai 400094, India}

	%\footnotetext[4]{jointly conceived the idea and supervised the project.}

%--------------------------------------------------------%
%            Abstract
%--------------------------------------------------------%
	\begin{abstract}
We propose a tunable platform for realizing two-dimensional second-order topological superconductivity (SOTSC) based on a quantum spin Hall insulator–superconductor heterostructure coupled to noncollinear magnetic textures. The system consists of a $d$-wave superconducting substrate, a quantum spin Hall layer, and a magnetic layer hosting meron–antimeron pairs or isolated merons and antimerons. Treating the superconducting order self-consistently within an extended Hubbard model and solving the Bogoliubov–de Gennes equations, we show that the $d$-wave pairing remains finite, although spatially modulated, over a broad range of exchange coupling, staggered mass, and spin–orbit coupling. For a meron–antimeron pair lattice, the quasiparticle spectrum exhibits two exchange-driven gap closings that separate three phases: a SOTSC phase with eight Majorana corner modes (MCMs), a second SOTSC phase with four MCMs, and a topologically trivial gapped superconducting phase.
Isolated merons display qualitatively similar higher-order topological phases. An effective low-energy edge theory attributes the tunability of MCMs number to the competition between exchange coupling and $d$-wave superconducting order induced edge masses, whose relative strengths are modified self-consistently by the magnetic texture. For comparison, a proximitized $s$-wave superconductor supports four MCMs but does not exhibit a tunable change in their number. Our results establish meronic spin textures as a route toward controlling multiple MCMs in hybrid quantum materials.

	%	We demonstrate that topological magnetic spin textures, including meron-antimeron pairs, vortex-antivortex pairs, and isolated merons/antimerons, can be integrated with $d$-wave superconductivity in a quantum spin Hall insulator (QSHI) heterostructure to engineer tunable higher-order topological phases. Through self-consistent solution of the superconducting gap equations in a real-space tight-binding framework, we establish the coexistence of $d$-wave pairing with these complex spin textures and map the parameter regimes supporting this pairing. For meron-antimeron spin textures, we discover a second-order topological superconducting (SOTSC) phase in symmetry class D hosting Majorana corner modes (MCMs), with a parameter-tunable transition between eight and four MCMs controlled by exchange coupling strength. We further examined the isolated meron and antimeron textures, for which a similar SOTSC phase persisted. When the out-of-plane spin components vanish, the texture reduces to a vortex–antivortex configuration, and the symmetry class changes to the BDI. The corresponding tunable SOTSC phase remains, and the bulk topology is characterized by the winding number of the quadrupole moment.In contrast, $s$-wave proximitized systems host four MCMs without tunable transitions, highlighting the unique role of $d$-wave pairing symmetry in enabling topological tunability. Our results show that the superconducting order parameter persists in the presence of these topological spin textures, enabling the engineering of a higher-order topology and a tunable number of MCMs.
	\end{abstract}
%--------------------------------------------------------
%--------------------------------------------------------

\maketitle

\section{Introduction}
	
Noncollinear magnetic textures characterized by topological charges, spanning both fractional and integer values--for \eg $\Nsk=\pm 1/2$ for merons and antimerons and $\Nsk=\pm 1$ for skyrmions and anti-skyrmions--offer a promising platform to manipulating quantum states in  %topological heterostructures.  
magnetic heterostructures and to spintronic applications~\cite{Nagaosa2013, Yu2012, Koshibae2016, Yu2018, Jani2021, Muhlbauer_2009, Bhukta2024,Nandy_PRL2016,Banik_AdvSci2025,Banik_PRBLett2026}.
%    This control arises from the spatially varying     exchange fields that these textures induce in adjacent materials. %Such textures 
Merons and antimerons have been observed and stabilized near room temperature in chiral magnets, antiferromagnetic systems, and synthetic magnetic platforms~\cite{Yu2018,Jani2021,Bhukta2024,Nagaosa2013}.
Their characteristic size, helicity, polarity, and stability can be tuned by engineering the magnetic anisotropy and exchange interactions, applying external perturbations, and exploiting interfacial Dzyaloshinskii–Moriya interactions~\cite{Heo_SciRep2016, Radaelli_2020, Conte_NanoLett2020, Jani2022}. 
Because the local magnetization varies continuously in space, these textures generate spatially modulated exchange fields in adjacent materials.
These properties enable topological spin textures attractive not only for spintronics application but also for designing spatially structured quantum states 
in magnet-superconductor hybrid (MSH) heterostructures~\cite{LoConte2024}.
%    Their ability to control adjacent quantum states stems from the spatially varying exchange fields they generate. 
    %Such textures have been mostly observed in chiral magnets, antiferromagnetic systems, and synthetic magnetic platforms, including meron–antimeron and bimeron configurations realized at or near room temperature~\cite{Yu2018,Jani2021,Bhukta2024,Nagaosa2013,Yu2018}.
    
Such spatially structured quantum states are especially compelling in topological superconductors (TSCs), which can host Majorana zero modes (MZMs). These are 
exotic quasiparticle excitations with non-Abelian exchange statistics that are of potential relevance to fault-tolerant topological quantum computation~\cite{Kitaev2001,Cheng_2011,Ivanov2001,SDSarma2008,beenakker2013search,Alicea_2012,Leijnse_2012,Kitaev2009}. 
In a conventional first-order TSC, Majorana modes appear as gapless boundary excitations localized on the ($d-1$)-dimensional boundary of a $d$-dimensional system. 
Higher-order topological superconductors (HOTSCs), in contrast, extend this bulk–boundary correspondence by supporting gapless Majorana states on boundaries of dimension lower than $d-1$; more generally, an $n^{\rm{th}}$-order TSC can host such states on $(d-n)$-dimensional boundaries~\cite{Geier_2018,Zhong_2017,Khalaf_2018,zhong2018,Volovik2010}. In particular, a two-dimensional (2D) second-order topological superconductor (SOTSC) can support zero-dimensional Majorana corner modes (MCMs) localized at the intersections of adjacent gapped edges.
This motivates the study of MSH systems in which spatially modulated exchange fields can whether engineer the edge-mass structure and thereby control the emergence, number, and localization of MCMs.

Over the past decade, engineering TSCs hosting Majorana modes have attracted considerable attention. A paradigmatic platform employs one-dimensional (1D) semiconductor nanowires in which Rashba spin–orbit coupling (SOC), Zeeman splitting via an external magnetic field, and proximity-induced $s$-wave superconductivity combine to generate the TSC phase~\cite{Fu2008,Lutchyn2010,Oreg2010,Mourik2012,Das2012,Albrecht2016}. 
An alternative route employs arrays of magnetic adatoms deposited on conventional $s$-wave superconductors, where hybridized Yu–Shiba–Rusinov states~\cite{Yu1965,Shiba1968,Rusinov1969,Eigler1997} can form topological bands capable of supporting MZMs~\cite{Felix2013, AliYazdani2013, DanielLoss2013, PascalSimon2013, MFranz2013, Eugene2013}. 
This magnetic route to realize TSC phases can be naturally extended to 2D, where spatially structured topological order provides additional control over the superconducting quasiparticle spectrum and boundary states. In recent years, 2D MSH systems incorporating noncollinear magnetic textures have emerged as promising platforms for realizing both first-order and higher-order TSCs~\cite{Pritam1_2024,Pritam2_2024}. 
In particular, magnetic skyrmions~\cite{DLoss_2016, Balatsky22016, Asle_2023}, skyrmion lattices~\cite{Mascot2021, Zlotnikov2021}, deformable skyrmionic configurations~\cite{Garnier2019}, and noncollinear arrangements of magnetic adatoms coupled to conventional $s$-wave or unconventional $d$-wave superconductors 
have been predicted to support Majorana edge modes, MCMs, and Majorana bound states localized near skyrmion or vortex cores~\cite{DLoss_2016,Mascot2021,Li_2026,Garnier2019,Neupert2016,Minakshi2025,Pritam1_2024,Pritam2_2024}.
Within the broader context of HOTSCs, several theoretical proposals have explored quantum spin Hall insulator (QSHI)--superconductor heterostructures~\cite{zhong2018,Cheng-Cheng_2018,Franco_2018}, bicollinear antiferromagnets~\cite{DasSharma_2019}, $\pi$-junction Rashba layers~\cite{Dloss_2019}, altermagnet heterostructures~\cite{Cheng-Cheng_2023}, unconventional $p$-wave magnets~\cite{Das2026}, noncollinear magnetic textures in proximity to conventional $s$-wave superconductors~\cite{Pritam1_2024,Pritam2_2024}, and unconventional $d$-wave pairing~\cite{zhong2018,Cheng-Cheng_2018,Minakshi2025}.
Nevertheless, the realization and control of HOTSC phases in 2D MSH systems through fractional topological spin textures (meron, antimeron, and bound meron--antimeron pair configurations) remain largely unexplored. 
In particular, their effects associated with unconventional superconducting order have not been systematically established, especially when the superconducting order parameter is treated self-consistently rather than as a fixed pairing amplitude. 

In this work, we investigate a three-layer heterostructure comprising of a $d$-wave superconductor substrate, a QSHI layer in the middle, and a magnetic layer on top that hosts noncollinear meronic textures.
Superconductivity is treated self-consistently using an extended Hubbard interaction: an attractive nearest-neighbor interaction produces $d$-wave pairing, whereas an attractive on-site interaction generates $s$-wave pairing.
By solving the Bogoliubov–de Gennes
(BdG) equations self-consistently, we obtain the spatial profile of the superconducting order parameter and analyze its modulation induced by the underlying spin textures.
We first consider meron–antimeron pairs and subsequently examine the effect of isolated merons and antimerons. In presence of these textures, the superconducting order remains finite but becomes spatially nonuniform. For the meron–antimeron configuration, the system supports SOTSC phases with either eight or four MCMs, separated by exchange-coupling-driven topological transitions; isolated merons exhibit qualitatively similar behavior.
The results confirm that the SOTSC phases with tunable MCMs, reported for fixed pairing amplitudes~\cite{Minakshi2025} persist when the $d$-wave order is obtained self-consistently. The complementary $s$-wave case, treated in Appendix~\ref{App_C} self-consistently for an attractive on-site Hubbard interaction, supports four MCMs but does not exhibit any tunable criterion in their number.

The remainder of this paper is organized as follows. Section~\ref{Sec:II} introduces the tight-binding model and the self-consistent framework used to determine the superconducting order parameter. Our analysis of meron- and antimeron-based magnetic textures is presented in Secs.~\ref{Sec:III} and~\ref{Sec:IV}.  We begin with meron–antimeron pair configurations in Sec.~\ref{Sec:III}, focusing on their topological phases and associated MCMs in the hybrid structure. 
We subsequently turn to isolated merons and antimerons in Sec.~\ref{Sec:IV}, where we examine the emergence of SOTSC phases and MCMs.
Finally, Sec.~\ref{Sec:V} summarizes our main findings and presents our conclusions.
In Appendix~\ref{App_A}, we numerically verify the topological charges ($\Nsk$) of the spin textures considered. 
In Appendix~\ref{App_SC}, we derive the mean-field BdG Hamiltonian and describe the numerical procedure used to determine the pairing self-consistently.
Appendix~\ref{App_B} investigates whether these phases persist in other geometries like circular and triangular geometries.  
Appendix~\ref{App_C} presents results for a QSHI layer coupled to meron--antimeron textures in close proximity to a self-consistently determined $s$-wave superconductor. Finally, in Appendix~\ref{App_D}, we analyze the low-energy theory to understand the origin of the MCMs and their tunability.
%======================================================
\section{Model and  Method}\label{Sec:II}
%======================================================
	\begin{figure}[t]
		\centering
		\includegraphics[width=\columnwidth]{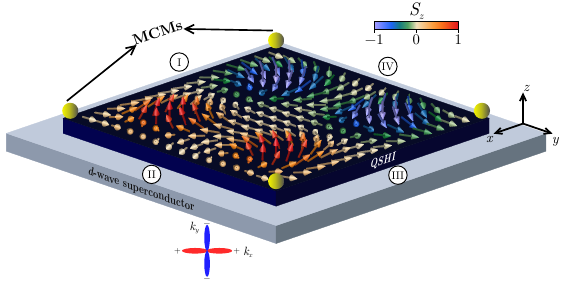}
		\caption{Schematic illustration of the three-layer heterostructure considered in this work. The bottom layer is a $d$-wave superconductor, the middle layer is a quantum spin Hall insulator (QSHI), and the top layer hosts two meron--antimeron pairs with texture centers located at the four quadrant centers of the lattice, with $q_i$ and $p_i$ as defined in Sec.~\ref{Sec:II} and core radius $\rc=4$. The arrows represent the spin vectors, and their color represents the out-of-plane spin component $S_z$. The yellow spheres at the corners denote the Majorana corner modes (MCMs), I--IV label the four edges, and the four-lobed structure in the $(k_x, k_y)$ plane depicts the $d_{x^2-y^2}$-wave pairing symmetry of the superconductor.
		}
		\label{fig:Fig1}
	\end{figure}
%=============================================================
We consider the tight-binding lattice model of a heterostructure consisting of a 2D QSHI on a $d$-wave superconducting substrate, coupled to a layer of topological magnetic textures on top, as schematically illustrated in Fig.~\ref{fig:Fig1}. 
The corresponding Hamiltonian is given by
	\begin{equation}
		H=H_0+H_{\mathrm{int}}\ .
	\end{equation}
Here, $H_0$ denotes the real-space lattice Hamiltonian of the QSHI coupled to the topological magnetic textures, whereas $H_{\mathrm{int}}$ describes the extended Hubbard interaction responsible for generating the superconducting pairing. The noninteracting Hamiltonian $H_0$ takes the form~\cite{Pritam2_2024, Minakshi2025}
%\textcolor{red}{
	\begin{equation}
		\begin{aligned}
			H_0
			=
			&
			\sum_{ \mathbf{r}_i,\mathbf{r}_j }
			c^\dagger_{\mathbf{r}_i}
			\,
			h_{\mathbf{r}_i,\mathbf{r}_j}
			\,
			c_{\mathbf{r}_j}
			\\
			=
			&
			\sum_{\mathbf{r}_i}
			c_{\mathbf{r}_i}^{\dagger}
			\bigg[
			\left\{
			\epsilon_0 \Gamma_1
			+
			J \mathbf{S}(\mathbf{r}_i)\cdot \mathbf{\Gamma}
			\right\}
			c_{\mathbf{r}_i}
			-
			\left(
			t\Gamma_1+i\lambda_x\Gamma_5
			\right)
			c_{\mathbf{r}_i+\hat{x}}
                \\
                &
                -
			\left(
			t\Gamma_1+i\lambda_y\Gamma_6
			\right)
			c_{\mathbf{r}_i + \hat{y}}
			\bigg]
			+\mathrm{H.c.}\ ,
			\label{eq:H_0}
		\end{aligned}
	\end{equation} 
where we consider a square lattice with coordinates denoted by $\mathbf r_i = (x_i, y_i)$, and $\hat x$ ($\hat y$) denotes the unit vector along the $x$ ($y$) direction. 
Throughout this paper, we set the lattice constant to $1$. 
At each site, the basis is defined in combined orbital-spin space as $c_{\mathbf{r}_i}=\left(c_{a\uparrow},c_{a\downarrow},c_{b\uparrow},c_{b\downarrow}\right)_{\mathbf r_i}^T$. 
Here, $h_{\mathbf r_i, \mathbf r_j}$ is the $4 \times 4$ block of the normal-state Hamiltonian, nonzero only for on-site and nearest-neighbor (NN) sites, and $\mathbf \Gamma \equiv \{\Gamma_2, \Gamma_3, \Gamma_4\}$.
The $\Gamma$ matrices are defined as
	\begin{align*}
		\Gamma_1 &= \sigma_zs_0, & \Gamma_2 &= \sigma_0s_x, & \Gamma_3 &= \sigma_0s_y, \\
		\Gamma_4 &= \sigma_0s_z, & \Gamma_5 &= \sigma_xs_z, & \Gamma_6 &=\sigma_y s_0.
	\end{align*}
Here, $\sigma_n$ and $s_n$ ($n=x,y,z$) are Pauli matrices acting in the orbital and spin spaces, respectively, while $\sigma_0$ and $s_0$ denote the corresponding identity matrices.  
%where $h_{\mathbf r_i,\mathbf r_j}$ denotes the $4 \times 4$ block of the normal-state Hamiltonian in orbital--spin space.
%The lattice position is specified by $\mathbf{r} = (l,m)$, where $l$ and $m$ are dimensionless integer coordinates measured in units of the lattice vectors along $x$ and $y$ directions.
%The operators are expressed in the combined orbital–spin basis as $c_{\mathbf{r}_i} \equiv c_i=\left(c_{i,a\uparrow},c_{i,a\downarrow},c_{i,b\uparrow},c_{i,b\downarrow}\right)^T$ or explicitely, $c_{\mathbf r}\equiv c_{l,m}=\left(c_{l,m,a\uparrow},c_{l,m,a\downarrow},c_{l,m,b\uparrow},c_{l,m,b\downarrow}\right)^T$.
The parameters $\epsilon_0$, $t$, $\lambda_x$, $\lambda_y$, and $J$ represent the staggered mass term, NN hopping amplitude, SOC strengths along the $x$ and $y$ directions, and local exchange coupling between the topological spin texture and the QSHI electrons, respectively. For $J=0$, $H_0$ reduces to the Bernevig--Hughes--Zhang (BHZ) model of a 2D QSHI~\cite{BHZ_2006,BHZ_2007}.
%In the combined orbital and spin basis $c_{m}=\left(c_{ma\uparrow},c_{ma\downarrow},c_{mb\uparrow},c_{mb\downarrow}\right)^T$, where $a,b$ label the orbital degrees of freedom, $\uparrow,\downarrow$ label the spin indices, and $T$ denotes the transpose operation. 
The magnetic textures considered in this work, which include isolated merons, isolated antimerons, and meron–antimeron pair lattice configurations, are modeled using a local classical spin field $\mathbf{S(r)}$ of constant magnitude $S$.
The spin field at each site is then given by
\begin{equation}
	\mathbf{S}(\mathbf{r}) = S 
	\begin{pmatrix}
		\sin f(\mathbf{r}) \cos[\phi(\mathbf{r}) + \gamma] \\[0.3em]
		\sin f(\mathbf{r}) \sin[\phi(\mathbf{r}) + \gamma] \\[0.3em]
		\cos f(\mathbf{r})
	\end{pmatrix}^T ,
	\label{Eq_Spin_Field}
\end{equation}
%	\begin{equation}
%		\begin{aligned}
%		\mathbf{S}(\mathbf{r}) = \\ 
%       & S \bigl(\sin f(\mathbf{r}) \cos[\phi(\mathbf{r}) 
%		+ \gamma],\, \sin f(\mathbf{r}) \sin[\phi(\mathbf{r}) + \gamma],\,
%		\cos f(\mathbf{r}) \bigr),
%        &
%		\label{Eq_Spin_Field}
%        \end{aligned}
%	\end{equation}
where $f(\mathbf{r})$ and $\phi(\mathbf{r})$ denote the polar and azimuthal angles, respectively, and $\gamma$ is the helicity parameter, fixed to $\gamma = \pi/2$ throughout this work; we also set $S=1$.
	
For an isolated texture centered at $\mathbf{R}_0=(X_0,Y_0)$, the azimuthal and polar angle profiles are parametrized by two integers $(q, p)$, where $q = \pm 1$ controls the azimuthal winding of the in-plane spin components and $p = \pm 1$ controls the orientation of the out-of-plane spin component at the texture core. 
Specifically, the azimuthal angle is chosen as
	\begin{equation}
		\phi(\mathbf{r})= q\,
		\tan^{-1}\!\left(\frac{y-Y_0}{x-X_0}\right),
		\label{Eq_Azim_Isolated}
	\end{equation}
	while the polar angle profile is given by
	\begin{equation}
		f(\mathbf{r})=
		\frac{\pi}{2}
		\left[1 + p\,
		\exp\!\left(-\frac{|\mathbf{r}-\mathbf{R}_0|^2}{\rc^2}\right)
		\right],
		\label{Eq_Polar_Isolated}
	\end{equation}
where $\rc$ is the core radius, which determines the spatial extent of the out-of-plane spin 
component. Here, $\tan^{-1}$ denotes the four-quadrant arctangent, so that $\phi$ winds by $2\pi q$ around the core. 
The four possible combinations of $(q, p)$ give rise to distinct texture configurations with different topological charges, as detailed in Table~\ref{tab:textures}. 
	\begin{table}[h]
		\centering
		\begin{tabular}{cccc}
			\hline\hline
			$q$ & $p$ & Core $S_z$ & $\Nsk$ \\
			\hline
			$+1$ & $+1$ & $-1$ (down) & $-1/2$ (meron)     \\
			$+1$ & $-1$ & $+1$ (up)   & $+1/2$ (antimeron) \\
			$-1$ & $+1$ & $-1$ (down) & $+1/2$ (antimeron) \\
			$-1$ & $-1$ & $+1$ (up)   & $-1/2$ (meron)     \\
			\hline\hline
		\end{tabular}
		\caption{Classification of isolated spin textures carrying topological charge $\Nsk$. This charge is defined according to winding number $q$ and core orientation $p$; for the profiles of Eqs.~\eqref{Eq_Azim_Isolated} and~\eqref{Eq_Polar_Isolated}, $\Nsk=-qp/2$.}
		\label{tab:textures}
	\end{table}
Throughout this work, we adopt the convention that a meron (antimeron) carries topological charge $\Nsk = -1/2$ ($+1/2$)~\cite{Yu2018}.
	
A lattice configuration of meron--antimeron pairs is obtained by arranging four such isolated textures with topological charges of opposite sign in the lower and upper rows, such that the total charge vanishes. 
This meron--antimeron pair lattice, which is the configuration used throughout this work and is shown in Fig.~\ref{fig:Fig1}, is constructed by fixing the cores at $\mathbf{R}_1=\left(\tfrac{N_x}{4},\tfrac{N_y}{4}\right)$, $\mathbf{R}_2=\left(\tfrac{3N_x}{4},\tfrac{N_y}{4}\right)$, $\mathbf{R}_3=\left(\tfrac{N_x}{4},\tfrac{3N_y}{4}\right)$, and $\mathbf{R}_4=\left(\tfrac{3N_x}{4},\tfrac{3N_y}{4}\right)$, where $N_x$ ($N_y$) denotes the number of lattice sites along the $x$ ($y$) direction.
We choose winding numbers $(q_1,q_2,q_3,q_4)=(+1,-1,-1,+1)$ and core orientation parameters $(p_1,p_2,p_3,p_4)=(+1,-1,+1,-1)$, which gives merons ($\Nsk=-1/2$) at $\mathbf{R}_1$ and $\mathbf{R}_2$ and antimerons ($\Nsk=+1/2$) at $\mathbf{R}_3$ and $\mathbf{R}_4$.
The azimuthal and polar angle profiles of this lattice are then given by
\begin{equation}
    \phi(\mathbf{r})=
    \sum_{i=1}^{4} q_i\,
    \tan^{-1}\!\left(\frac{y-Y_i}{x-X_i}\right),
    \label{Eq_Azim_MAM}
\end{equation}
\begin{equation}
    f(\mathbf{r})=
    \frac{\pi}{2}
    \sum_{i=1}^{4} p_i\,
    \exp\!\left(-\frac{|\mathbf{r}-\mathbf{R}_i|^2}{\rc^2}\right)
    +\frac{\pi}{2},
    \label{Eq_Polar_MAM}
\end{equation}
where $\mathbf{R}_i=(X_i,Y_i)$.
Details of the topological charges and winding numbers associated with these textures are provided in Appendix~\ref{App_A}.
	
To incorporate superconductivity, we model the pairing as an intrinsic lattice attraction, as in Ref.~\cite{De2023} for the proximity effect in a QSHI, and include the extended Hubbard interaction~\cite{Micnas_1990, Zhu2016BdG}:
	\begin{equation}
		H_{\mathrm{int}}
		= U \sum_{\mathbf r_i,\alpha} n_{\mathbf r_i\alpha\uparrow} n_{\mathbf r_i\alpha \downarrow}
		-\frac{V}{2}
		\sum_{\substack{\langle \mathbf r_i,\mathbf r_j\rangle,\\ 
        \alpha,\sigma,\sigma'}}
		n_{\mathbf r_i\alpha\sigma}
		n_{\mathbf r_j\alpha\sigma'},
			\label{Eq_Hint}
	\end{equation}
where $n_{\mathbf r_i\alpha\sigma} = c^\dagger_{\mathbf r_i\alpha\sigma} c_{\mathbf r_i\alpha\sigma}$ is the electron number operator at the lattice site $\textbf{r}_i$ with orbital $\alpha$ and spin $\sigma$, and $\langle \mathbf r_i,\mathbf r_j\rangle$ runs over NN pairs of sites, counting both orientations $(\mathbf r_i,\mathbf r_j)$ and $(\mathbf r_j,\mathbf r_i)$.
Here, $U$ and $V$ are the on-site and NN Hubbard interaction strengths, respectively. 
An attractive on-site interaction ($U<0$) drives $s$-wave pairing~\cite{N_Stefanakis_2003, Zhu2016BdG}, whereas an attractive NN interaction $V>0$ is known to stabilize a $d$-wave superconducting state in related lattice models~\cite{Micnas_1990,N_Stefanakis_2003}.
Using a self-consistent formulation, we study the $d$-wave case $(U=0,\, V > 0)$ in the main text, while the $s$-wave case $(U<0,\, V = 0)$ is presented in Appendix~\ref{App_C}.
	
Upon performing a standard mean-field decoupling of the interaction in the superconducting pairing channel and setting $U=0$ and $V>0$, we arrive at 
	\begin{equation}
		\begin{aligned}
			H_{\mathrm{MF}}
			=
			&\sum_{\mathbf r_i,\mathbf r_j}
			c^{\dagger}_{\mathbf r_i}
			\,
			h^{\phantom{\dagger}}_{\mathbf r_i,\mathbf r_j}
			\,
			c^{\phantom{\dagger}}_{\mathbf r_j}
			\\
			&
			+\sum_{\langle \mathbf r_i , \mathbf r_j\rangle,\alpha}
			\left(
			\Delta (\mathbf r_i, \mathbf r_j )
			c^{\dagger}_{i\alpha\uparrow}
			c^{\dagger}_{j\alpha\downarrow}
			+ H.c.
			\right)
			\\
			&
			+\sum_{\langle \mathbf r_i, \mathbf r_j \rangle}
			\frac{4 |\Delta (\mathbf r_i, \mathbf r_j )|^2}{V}.
		\end{aligned}\label{Eq_MF}
	\end{equation}
The superconducting order parameter defined on NN bonds is given by~\cite{Zhu2016BdG}
	\begin{equation}
		\Delta (\mathbf r_i, \mathbf r_j )
		=
		\frac{V}{2}
		\left[
		\langle
		c_{\mathbf r_i\alpha\uparrow}
		c_{\mathbf r_j\alpha\downarrow}
		\rangle
		-
		\langle
		c_{\mathbf r_i\alpha\downarrow}
		c_{\mathbf r_j\alpha\uparrow}
		\rangle
		\right]. 
	\end{equation}
For simplicity, the pairing amplitude is assumed to be identical in both orbital sectors ($\alpha \in \{a,b\}$). 
%In the bulk, $\Delta_{ij}$ is positive for NN bonds oriented along the crystallographic $\hat{x}$ direction and negative for bonds oriented along the $\hat{y}$ direction.
Introducing the Nambu spinor 
	$\Psi_{\mathbf r_i} = (c_{a\uparrow}, c_{a\downarrow}, c_{b\uparrow}, c_{b\downarrow}, 
	c^\dagger_{a\uparrow}, c^\dagger_{a\downarrow}, c^\dagger_{b\uparrow}, 
	c^\dagger_{b\downarrow})_{\mathbf r_i}^T$, 
the mean-field Bogoliubov--de Gennes (BdG) Hamiltonian takes the form
	\begin{equation}
		H_{\mathrm{BdG}}^{\mathrm{MF}}
		=
		\frac{1}{2}
		\sum_{\mathbf r_i,\mathbf r_j}
		\Psi_{\mathbf r_i}^\dagger
		\,
		\mathcal H_{\mathrm{BdG}}(\mathbf r_i, \mathbf r_j)
		\,
		\Psi_{\mathbf r_j}
		+
		\sum_{\langle \mathbf r_i,\mathbf r_j\rangle}
		\frac{4|\Delta (\mathbf r_i, \mathbf r_j )|^2}{V},
	\end{equation}
where
	\begin{equation}
		\mathcal H_{\mathrm{BdG}}(\mathbf r_i, \mathbf r_j)
		=
		\begin{pmatrix}
			h_{\mathbf r_i, \mathbf r_j}
			&
			\Delta (\mathbf r_i, \mathbf r_j )\,(\sigma_0 \, is_y)
			\\[4pt]
			\Delta (\mathbf r_i, \mathbf r_j )^{*}\,(\sigma_0 \, is_y)^{\dagger}
			&
			-h^{*}_{\mathbf r_i , \mathbf r_j}
		\end{pmatrix}.
		\label{Eq_BdG_ham}
	\end{equation}
Here, $\mathcal{H}_{\mathrm{BdG}}(\mathbf r_i, \mathbf r_j)$ is an $8 \times 8$ matrix spanning the particle-hole, orbital and spin degrees of freedom.
A Bogoliubov transformation therefore leads to the following BdG equations~\cite{Zhu2016BdG,deGennes1999,Wennerdal2020,N_Stefanakis_2003} (see Appendix~\ref{App_SC} for the derivation)
	\begin{equation}
		\sum_{\mathbf r_j}
		\mathcal H_{\mathrm{BdG}}(\mathbf r_i , \mathbf r_j)
		\begin{pmatrix}
			u^{n}(\mathbf r_j) \\
			v^{n}(\mathbf r_j)
		\end{pmatrix}
		=
		E_n
		\begin{pmatrix}
			u^{n}(\mathbf r_i) \\
			v^{n}(\mathbf r_i)
		\end{pmatrix},
		\label{Eq_BdG}
	\end{equation}
where $E_n$ is the quasiparticle eigenenergy and $(u^n(\mathbf r_i), v^n(\mathbf r_i))^T$ is the corresponding eight-component Bogoliubov eigenvector, with $u^n (\mathbf r_i) = (u^n_{a\uparrow}, u^n_{a\downarrow}, 
u^n_{b\uparrow}, u^n_{b\downarrow})(\mathbf r_i)$ and similarly for $v^n(\mathbf r_i)$. 
Both positive- and negative-energy eigenstates are retained, with the factor of $\tfrac{1}{2}$ in $H_{\mathrm{BdG}}^{\mathrm{MF}}$ accounting for the particle-hole redundancy of the Nambu representation.
This set of BdG equations is subject to the self-consistency conditions on the bond-dependent order parameter~\cite{Zhu2016BdG}
\begin{equation}
	\begin{aligned}
		\Delta (\mathbf r_i, \mathbf r_j )
		= &
		\frac{V}{4}
		\sum_n 
        \Big[
		u^{n}_{\uparrow}(\mathbf r_i)
		v^{n*}_{\downarrow}(\mathbf r_j)
		\\ 
        & 
        + u^{n}_{\uparrow}(\mathbf r_j)
		v^{n*}_{\downarrow}(\mathbf r_i)
		\Big] 
		\tanh\!\left(
		\frac{E_n}{2T}
		\right).
		\label{Eq_GapEq}
	\end{aligned}
\end{equation}
Here, we drop the orbital index; the components $u^n_{\sigma}(\mathbf r_i)$, $v^n_{\sigma}(\mathbf r_i)$ refer to a single orbital sector. We solve the gap equation~(\ref{Eq_GapEq}) self-consistently, iterating until $\Delta$ converges within a prescribed tolerance ($10^{-6}$); the numerical procedure, which iterates from randomized initial conditions as in Ref.~\cite{De2023}, is described in Appendix~\ref{App_SC}.  All calculations are performed at zero temperature, $T = 0$, where $\tanh(E_n/2T)\to\mathrm{sgn}(E_n)$ in Eq.~(\ref{Eq_GapEq}). We adopt units in which $\hbar = k_B = e = 1$, and we set the NN hopping amplitude $t=1$ as the unit of energy.
The spatial profile of the superconducting order parameter is characterized by the local $d$-wave amplitude at site $\mathbf r_i$, defined as~\cite{N_Stefanakis_2003}

\begin{equation}
	\begin{aligned}
		\Delta_0(\mathbf r_i)
		=
		\frac{1}{4}
		\Big[
		&\Delta(\mathbf r_i,\mathbf r_i+\hat x)
		+\Delta(\mathbf r_i,\mathbf r_i-\hat x)
		\\
		&-\Delta(\mathbf r_i,\mathbf r_i+\hat y)
		-\Delta(\mathbf r_i,\mathbf r_i-\hat y)
		\Big],
	\end{aligned}
\end{equation}
which satisfies $\Delta (\mathbf r_i,\mathbf r_i+\hat x) = -\Delta(\mathbf r_i,\mathbf r_i+\hat y)$ in the bulk, consistent with $d_{x^2-y^2}$-wave pairing symmetry.
		
\begin{figure*}[t]
	\centering
		\includegraphics[width=\textwidth]{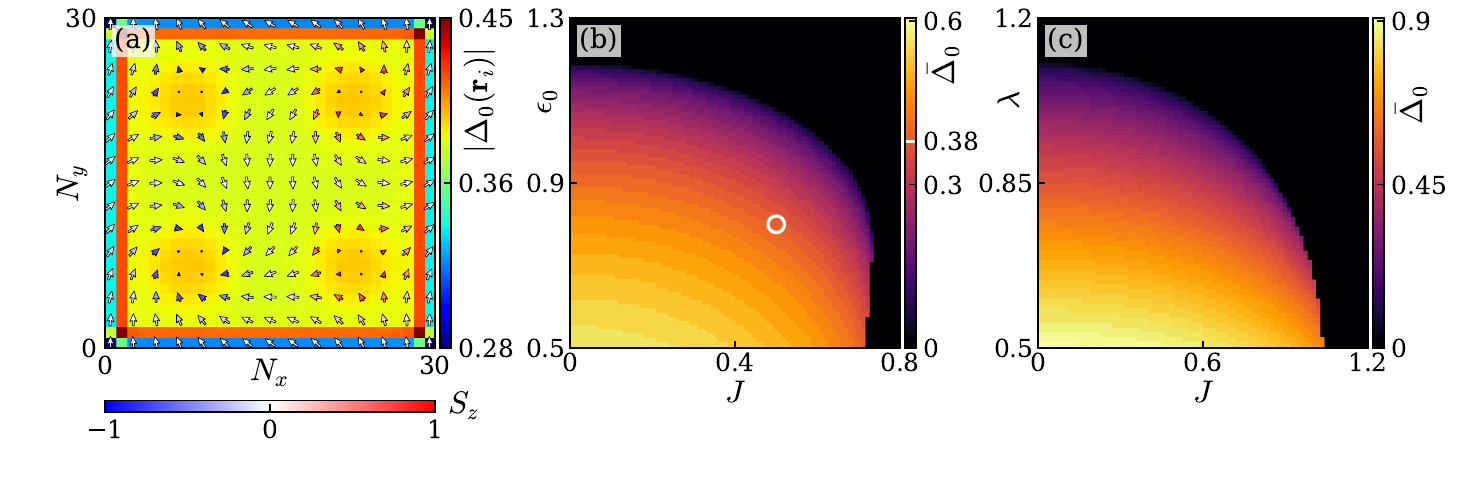}
		\caption{(a) Real-space distribution of the local $d$-wave order parameter $|\Delta_0(\mathbf{r}_i)|$ obtained self-consistently on a $30 \times 30$ lattice with OBC for the meron--antimeron spin texture at $J = 0.5t$, $\lambda = t$, and $\epsilon_0 = 0.8t$, for which $\bar{\Delta}_0=0.38t$. The arrows show the in-plane spin components of the texture, and their color represents $S_z$. (b) Spatially averaged superconducting order parameter $\bar{\Delta}_0$ [defined in Eq.~(\ref{Eq_avg_order_param})] in the $\epsilon_0$--$J$ plane at $\lambda = t$. The circle marks the parameters of panel (a). (c) $\bar{\Delta}_0$ in the $\lambda$--$J$ plane at $\epsilon_0 = t$. Throughout this calculation, we take isotropic SOC $\lambda_x = \lambda_y \equiv \lambda$, with $V=3.5t$, $\rc=4$, $\gamma = \pi/2$ and $t = 1$. }
		\label{fig:Fig.2}
\end{figure*}
%

%In the absence of the exchange coupling ($J=0$), the Hamiltonian in Eq.~\eqref{eq:H_0} reduces to the BHZ model describing a 2D QSHI ~\cite{BHZ_2006,BHZ_2007}.
%The BHZ model was originally proposed for HgTe/CdTe quantum wells, where the intrinsic SOC is described by the $\lambda_x$ and $\lambda_y$ terms in Eq.~\eqref{eq:H_0}. The BHZ model exhibits a 2D topological insulating phase characterized by gapless helical edge states protected by time-reversal symmetry (TRS).
%In the presence of noncollinear magnetic textures, the introduction of superconducting pairing can drive the system into a SOTSC phase hosting MCMs.
%Previous studies based on phenomenologically introduced superconducting order parameter have shown that an $s$-wave superconducting pairing supports four MCMs~\cite{Pritam2_2024}, whereas a $d$-wave superconducting pairing can generate a tunable SOTSC phase with a transition between eight and four MCMs~\cite{Minakshi2025}.
%However, whether these phases persist when the superconducting order parameter is treated self-consistently remains an open question. Motivated by these earlier results, here we investigate whether a self-consistently generated $d$-wave superconducting order can persist in the presence of meron--antimeron spin textures and whether such a system can host a similar tunable number of MCMs.
%In Sec.~\ref{Sec:III}, we present the results for the meron--antimeron pair configuration.

\section{Results}
\subsection{Meron--Antimeron Pairs}{\label{Sec:III}}

To investigate the emergence of HOTSC, we first consider the meron--antimeron pair lattice of Fig.~\ref{fig:Fig1}, in which the local spin vector is defined by Eq.~\eqref{Eq_Spin_Field}, while its spatial variation is encoded in the polar and azimuthal angles given by Eqs.~\eqref{Eq_Polar_MAM} and~\eqref{Eq_Azim_MAM} with the values of $q_i$ and $p_i$ given in Sec.~\ref{Sec:II}.
We then obtain the order parameter $\Delta(\mathbf r_i,\mathbf r_j)$ on every nearest-neighbor bond self-consistently, as described in Sec.~\ref{Sec:II} and Appendix~\ref{App_SC}.
A central question is whether this $d$-wave superconducting order remains stable in the presence of the nonuniform spin texture.
Figure~\ref{fig:Fig.2}(a) shows the spatial profile of the local order parameter $|\Delta_0(\mathbf r_i)|$, together with the corresponding spin texture, obtained from a self-consistent calculation for the heterostructure on a $30\times30$ lattice with open boundary conditions (OBC).
In the bulk, $|\Delta_0|\approx 0.38t$, and for finite exchange coupling ($J$) and SOC ($\lambda$) it becomes spatially nonuniform, exhibiting a slight enhancement of about $4\%$ near the meron and antimeron cores.
This enhancement occurs predominantly in regions where the local exchange field is oriented out of the plane, with $S_z$ dominant and the in-plane components ($S_x,S_y$) suppressed.

This behavior suggests that the out-of-plane exchange field is less detrimental to $d$-wave superconductivity than its in-plane counterpart.
The microscopic origin of this correlation remains an interesting subject for further investigation.
Near the boundary, $|\Delta_0|$ differs from its bulk value. The outermost sites show a lower value ($\approx 0.33t$, and $\approx 0.28t$ at the corners), because only three of the four bonds in the definition of $\Delta_0(\mathbf r_i)$ exist at an edge site (two at a corner), and the missing bonds enter as zero. The bond amplitudes themselves are enhanced near the edge ($\approx 0.49t$ for the bond perpendicular to the edge and $\approx 0.43t$ along it, compared with $\approx 0.38t$ in the bulk). This gives the enhanced values on the sites next to the edge ($\approx 0.42t$, up to $\approx 0.45t$ close to the corners).

To determine the parameter regime in which the self-consistent $d$-wave superconducting order persists, we use the spatially averaged order-parameter magnitude,
	\begin{equation}
		\bar{\Delta}_0=\frac{1}{N_xN_y}\sum_{\mathbf r_i} |\Delta_0(\mathbf r_i)|,
		\label{Eq_avg_order_param}
	\end{equation}
where $N_xN_y$ denotes the total number of lattice sites.
The resulting dependence of $\bar{\Delta}_0$ on the $\epsilon_0$--$J$ and $\lambda$--$J$ parameter planes is presented in Figs.~\ref{fig:Fig.2}(b) and ~\ref{fig:Fig.2}(c), respectively.
%As evident from Fig.~\ref{fig:Fig.2}(b), the spatially averaged pairing amplitude remains finite over a broad range of $J$ and $\epsilon_0$. This result indicates that the $d$-wave superconducting state remains stable in the presence of the spatially varying exchange field generated by the magnetic texture, although its magnitude is modulated locally.
As seen in Figs.~\ref{fig:Fig.2}(b) and~\ref{fig:Fig.2}(c), the spatially averaged pairing amplitude $\bar{\Delta}_0$ remains finite over a broad range of $J$, $\epsilon_0$, and $\lambda$, indicating that the $d$-wave superconducting state is robust against the spatially varying exchange field generated by the meron--antimeron texture.
Furthermore, increasing either the staggered mass or the SOC strength $\lambda$ for a fixed $J$ suppresses the order parameter. Nevertheless, $\bar{\Delta}_0$ remains finite over a substantial region of parameter space, establishing that bulk $d$-wave superconductivity survives in the presence of the meron--antimeron texture.
The self-consistent superconducting state is destroyed when any one of the staggered mass, the SOC, and the exchange coupling becomes sufficiently strong, with the other two held fixed.
	
\begin{figure}[t]
	\centering
	\includegraphics[width=\columnwidth]{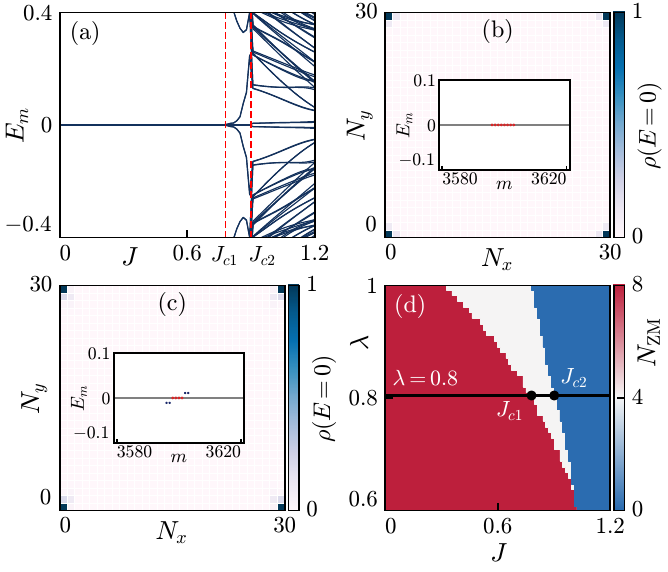}
		\caption{For the meron--antimeron pair configuration, (a) BdG eigenenergy spectrum as a function of $J$, showing two critical points $J_{c1}\approx0.83$ and $J_{c2}\approx0.90$ that separate the phases hosting eight and four MCMs. (b),(c) Spatially resolved LDOS at zero energy for $J = 0.3t$ and $J = 0.85t$, corresponding to the phases with eight and four MCMs, respectively, with insets showing the enlarged spectra around zero energy (states with $|E|<10^{-6}$ in red); each map is normalized to its maximum. (d) Number of
		zero-energy modes $N_{\mathrm{ZM}}$ in the $\lambda$--$J$ plane, which reveals three distinct phases: a trivial phase with no zero modes (blue), and topological phases hosting four (white) and eight
		(red) MCMs. Panels (a)--(c) use $\lambda = 0.8$, while $\lambda$ is varied
		along the vertical axis in panel (d), where the black line marks $\lambda=0.8$ and the dots indicate $J_{c1}$ and $J_{c2}$ shown in (a). Throughout this calculation, we set $\epsilon_0 = 1$, $t = 1$, $\rc = 2$, and $\gamma = \pi/2$, and the superconducting order parameter $\Delta_0$ is obtained self-consistently from the gap equation. All panels are computed for a $30\times30$ lattice with OBC.}
		\label{fig:Fig3}
\end{figure}

Having established the self-consistent $d$-wave superconducting state in the presence of the meron--antimeron texture (Fig.~\ref{fig:Fig1}), we next examine the associated low-energy quasiparticle excitations and their topological character.
To this end, we solve the self-consistent BdG equations in Eq.~\eqref{Eq_BdG} for a finite system with OBC imposed along both spatial directions. 
The resulting quasiparticle excitation spectrum as a function of $J$ is shown in Fig.~\ref{fig:Fig3}(a) for $\epsilon_0=t$.
Two distinct gap-closing points are observed, at $J_{c1}\approx0.83$ and $J_{c2}\approx0.90$ (dashed vertical lines), indicating successive topological phase transitions.
To characterize the spatial structure of the zero-energy states, we compute the site-resolved local density of states (LDOS) at energy $E=0$, defined as~\cite{Wennerdal2020}
	\begin{equation}
		\rho(\mathbf r_i,E)=
		-\frac{1}{\pi}\,\mathrm{Im}
		\sum_m
		\left[
		\frac{|u^m(\mathbf r_i)|^2}{E-E_m+i\eta}
		+
		\frac{|v^m(\mathbf r_i)|^2}{E+E_m+i\eta}
		\right],\label{Eq_LDOS}
	\end{equation}
where the sum runs over all eigenstates, $|u^m(\mathbf r_i)|^2=\sum_{\alpha,\sigma}|u^m_{\alpha\sigma}(\mathbf r_i)|^2$ (and likewise for $v^m$), and $\eta$ is a small broadening parameter.
The zero-energy LDOS profiles for the two distinct topological phases, corresponding to $J<J_{c1}$ and $J_{c1}<J<J_{c2}$ are shown in Figs.~\ref{fig:Fig3}(b) and~\ref{fig:Fig3}(c), respectively.

The quasiparticle eigenvalues $E_m$, plotted as a function of the state index $m$, are shown in the insets.
In both phases, the zero-energy LDOS is strongly localized near the four corners of the sample, providing a direct real-space signature of the MCMs and confirming the SOTSC phases~\cite{zhong2018}.
The two phases are distinguished by the number of zero-energy modes, $N_{\mathrm{ZM}}$: the phase for $J<J_{c1}$ hosts eight MCMs, whereas the intermediate phase, $J_{c1}<J<J_{c2}$, hosts four MCMs.
Finally, for $J>J_{c2}$, the system becomes a topologically trivial gapped superconductor without zero-energy modes.
To establish the stability of these phases over a broader parameter range, we calculate the number of zero-energy modes, $N_{\mathrm{ZM}}$, throughout the $\lambda-J$ parameter plane, as shown in Fig.~\ref{fig:Fig3}(d).
Here, $N_{\mathrm{ZM}}$ is obtained by counting the quasiparticle states satisfying $|E|<10^{-6}$.
The resulting phase diagram consists of three distinct regions. The red region corresponds to an SOTSC phase hosting eight MCMs, the white region denotes an SOTSC phase with four MCMs, and the blue region represents a topologically trivial gapped superconducting phase with no zero-energy modes.  The black line in Fig.~\ref{fig:Fig3}(d) marks $\lambda=0.8$, the value used in Figs.~\ref{fig:Fig3}(a)--\ref{fig:Fig3}(c), and the dots where it crosses the phase boundaries correspond to $J_{c1}$ and $J_{c2}$.
We further verify that the MCMs persist in disc and triangular geometries, as shown in Appendix~\ref{App_B}.

\subsection*{Topological Characterization}
To characterize the different topological phases identified from the quasiparticle excitation spectrum and the zero-mode counting, we now compute the corresponding bulk topological invariant. 
For the present SOTSC phase, the appropriate invariant is the quadrupolar winding number $N_{xy}$. 
Unlike the conventional bulk quadrupole moment, which is $\mathbb{Z}_2$-valued and therefore cannot distinguish phases hosting multiple MCMs at each corner~\cite{Benalcazar_2017}, the quadrupolar winding number takes integer values and provides a more refined characterization of the higher-order topological phases. 
In particular, $N_{xy}=1$ corresponds to a SOTSC phase hosting a single MCM at each corner, while $N_{xy}=2$ characterizes the phase with two MCMs localized at each corner of the 2D system.
	
For finite $\rc$, $S_z\neq0$, and no effective TRS operator can be constructed. Consequently, the chiral operator, which is the product of the TRS and particle--hole operators, is absent as well. Only the particle--hole symmetry, $\mathcal C=\tau_x\mathcal K$ with $\mathcal C^2=+1$, remains intact, where $\mathcal K$ is complex conjugation. According to the Altland--Zirnbauer classification~\cite{AltlandZirnbauer1997}, the system then belongs to symmetry class D, and the quadrupolar winding number is not well defined. 
In real space, in this class, no readily computable real-space invariant is available that resolves the number of MCMs localized at each corner. We therefore characterize the bulk topology in the $\rc \to 0$ limit, where the out-of-plane component vanishes $(S_z=0)$, the texture becomes purely in-plane, and the mean-field BdG Hamiltonian [Eq.~\eqref{Eq_BdG_ham}] possesses a chiral symmetry,
	$
	\mathcal S^{\dagger}
	\mathcal H_{\mathrm{BdG}}
	\mathcal S
	=
	-
	\mathcal H_{\mathrm{BdG}},
	$
	where the chiral symmetry operator is
	$
	\mathcal S
	=
	\tau_y \sigma_0 s_x.
	$
The system then belongs to class BDI~\cite{AltlandZirnbauer1997}, for which $N_{xy}$ is well defined.
To construct $N_{xy}$, it is convenient to work in the basis that diagonalizes the chiral symmetry operator. Under the unitary transformation $U_c$, the chiral operator takes the diagonal form
	$
	U_c
	\mathcal S
	U_c^\dagger
	=
	\mathrm{diag}(1,1,\dots,1,-1,-1,\dots,-1),
	$
	and the BdG Hamiltonian can be expressed in the block off-diagonal form
	\begin{equation}
		\tilde{\mathcal H}_{\mathrm{BdG}}
		=
		U_c
		\mathcal H_{\mathrm{BdG}}
		U_c^\dagger
		=
		\begin{pmatrix}
			0 & h \\
			h^\dagger & 0
		\end{pmatrix},
	\end{equation}
	where $h$ denotes the off-diagonal block of the transformed Hamiltonian.
	
The chiral symmetry decomposes the Hilbert space into two subspaces, denoted by $A$ and $B$, corresponding to the $+1$ and $-1$ eigenvalues of the chiral symmetry operator $\mathcal S$, respectively. The associated projection operators are given by
	$
	U_A^{\mathcal S}
	=
	\sum_{\alpha\in A}
	|\alpha\rangle\langle\alpha|
	$
	and
	$
	U_B^{\mathcal S}
	=
	\sum_{\beta\in B}
	|\beta\rangle\langle\beta|.
	$
We then perform a singular value decomposition (SVD) of the off-diagonal block $h$ as
	$
	h
	=
	U_A
	\Sigma
	U_B^\dagger,
	$
where $\Sigma$ is the diagonal matrix containing the singular values, while $U_A$ and $U_B$ are unitary matrices formed from the corresponding singular vectors. Using these quantities, the quadrupolar winding number is defined as~\cite{Benalcazar_2022, A_Pal_2025}
	\begin{equation}
		N_{xy}
		=
		\frac{1}{2\pi i}
		\mathrm{Tr}
		\ln
		\left(
		\bar Q_A
		\bar Q_B^\dagger
		\right),
	\end{equation}
where the sublattice quadrupole operators are given by
	$
	\bar Q_{A,B}
	=
	U_{A,B}^\dagger
	U_{A,B}^{\mathcal S}
	Q
	U_{A,B}^{\mathcal S}
	U_{A,B}.
	$
The quadrupole operator is defined as
	$
	Q
	=
	\exp
	\left(
	-\frac{2\pi i\,xy}{N_xN_y}
	\right),
	$
where $N_x$ and $N_y$ denote the number of lattice sites along the $x$ and $y$ directions, respectively. The invariant $N_{xy}$ takes discrete integer values and directly characterizes the number of MCMs localized at each corner of the system.
	
	\begin{figure}[t]
		\centering
		\includegraphics[width=\columnwidth]{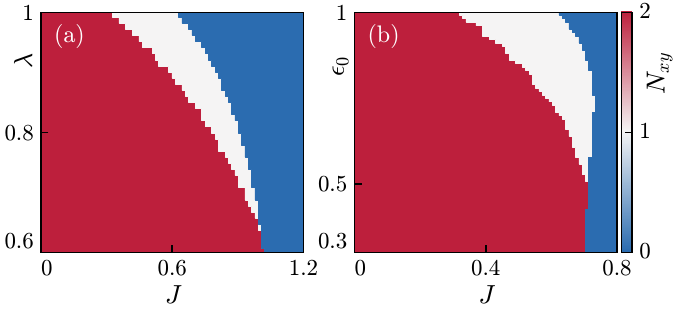}
		\caption{Quadrupole winding number $N_{xy}$ for the meron--antimeron pair configuration in the $\rc \to 0$ limit. (a) $N_{xy}$ in the $\lambda$--$J$ plane at $\epsilon_0 = 1$. (b) $N_{xy}$ in the $\epsilon_0$--$J$ plane at $\lambda = 1$. Both panels reveal three distinct phases: a trivial phase with $N_{xy} = 0$ (blue), a SOTSC phase with a single MCM per corner, $N_{xy} = 1$ (white), and a SOTSC phase with two MCMs per corner, $N_{xy} = 2$ (red). Throughout this calculation, we set $t = 1$ and $\gamma = \pi/2$, and the superconducting order parameter $\Delta$ is obtained self-consistently from the gap equation, with $V=3.5t$ as in Fig.~\ref{fig:Fig.2}.}
		\label{fig:Fig4}
	\end{figure}
	
In Fig.~\ref{fig:Fig4}, we summarize the topological characterization of the meron--antimeron pair configuration in the $\rc \to 0$ limit. 
Figure~\ref{fig:Fig4}(a) shows the quadrupolar winding number $N_{xy}$ obtained from the self-consistent BdG Hamiltonian in the $\lambda$--$J$ plane. 
The phase diagram exhibits three distinct regions: the red 
region with $N_{xy}=2$ corresponds to a SOTSC phase hosting two MCMs at each corner, the white region with $N_{xy}=1$ to a phase with a single MCM per corner, and the blue region with $N_{xy}=0$ to a topologically trivial gapped superconducting phase without MCMs. 
Figure~\ref{fig:Fig4}(b) presents $N_{xy}$ in the $\epsilon_0$--$J$ plane, where the same three phases appear. The phase boundaries qualitatively agree with the zero-mode counting of 
Fig.~\ref{fig:Fig3}(d), the $N_{xy}=2$ and $N_{xy}=1$ regions 
corresponding to the eight- and four-MCM phases, respectively. The quantitative differences arise because $N_{xy}$ is evaluated in the $\rc \to 0$ limit, whereas the zero-mode counting of Fig.~\ref{fig:Fig3}(d) is obtained at finite $\rc$.
In the limit $\rc \to 0$, we construct an effective low-energy model and edge theory to understand these phases, as detailed in Appendix~\ref{App_D}. 
In real space, one can also characterize the SOTSC phase in the chiral-symmetric BDI (for our system in the $\rc \to 0$ limit) and DIII classes~\cite{AltlandZirnbauer1997} using suitably constructed Bott indices~\cite{Luo_2025}.

\subsection{Isolated meron and antimeron configurations}\label{Sec:IV}

Having established that meron--antimeron pair textures can induce tunable SOTSC phases, we now ask whether these phases require a paired texture or can also arise from an isolated meron or antimeron. 
An isolated texture allows us to separate the effects that originate from the local structure of a single meron or antimeron from those that depend on the collective arrangement of multiple textures. 
%Although meron--antimeron pairs and bimeron configurations have been observed in several chiral and antiferromagnetic magnetic systems~\cite{Yu2018,Jani2021,Bhukta2024}, stabilizing isolated merons or antimerons generally requires additional mechanisms, such as interfacial Dzyaloshinskii--Moriya interactions, exchange engineering, or magnetic-anisotropy tuning~\cite{Radaelli_2020,Jani2022}. 

We first examine whether the self-consistent $d$-wave superconducting state persists in the presence of an isolated texture. 
The isolated meron and antimeron spin textures are described by Eq.~\eqref{Eq_Spin_Field}, with the azimuthal profile $\phi(\mathbf{r})$ given by Eq.~\eqref{Eq_Azim_Isolated} and the polar-angle profile $f(\mathbf{r})$ by Eq.~\eqref{Eq_Polar_Isolated}. 
The winding number $q$ and core orientation parameter $p$ distinguish four configurations: core-up and core-down merons, and core-up and core-down antimerons (Table~\ref{tab:textures} in Sec.~\ref{Sec:II}).

\begin{figure}[t]
		\centering
		\includegraphics[width=\columnwidth]{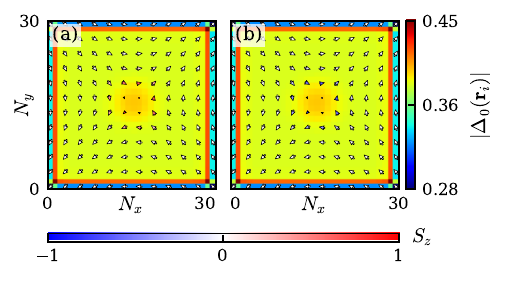}
		\caption{Real-space distribution of the local order parameter magnitude $|\Delta_0(\mathbf{r}_i)|$ obtained self-consistently on a $30 \times 30$ lattice with OBC for (a) an isolated core-up meron with $(q,p)=(-1,-1)$ and (b) an isolated core-down antimeron with $(q,p)=(-1,1)$, with the corresponding spin textures overlaid. The arrows show the in-plane spin components, and their color represents $S_z$. Throughout this calculation, we take $t=1$, $\epsilon_0=0.8t$, $J=0.5t$, $\lambda_x=\lambda_y=t$, $\rc=4$, $\gamma=\pi/2$, $U=0$ and $V=3.5t$.}
		\label{fig:Fig5}
	\end{figure}

Following the self-consistent procedure used for the meron--antimeron pair in Sec.~\ref{Sec:III} (see Appendix~\ref{App_SC}), we solve the gap equation~\eqref{Eq_GapEq} for an isolated core-up meron and an isolated core-down antimeron. 
Figure~\ref{fig:Fig5} shows the resulting spatial profiles of $|\Delta_0 (\mathbf r_i)|$ superimposed on the corresponding spin textures. 
In both cases, the superconducting order remains finite and becomes spatially nonuniform, with a slight enhancement near the texture core, as also observed for the meron--antimeron pair.
The magnitude $|\Delta_0|$ is identical for the meron and the antimeron: reversing $S_z$ is equivalent to time reversal combined with a $\pi$ spin rotation about the $z$ axis, which maps $\Delta$ to its complex conjugate for the singlet pairing considered here~\cite{Sigrist1991}.
These results show that the self-consistent $d$-wave state does not require a paired texture. 
Since the two configurations exhibit qualitatively similar superconducting behavior, we focus below on the isolated core-up meron.

\begin{figure}[htbp]
		\centering
		\includegraphics[width=\columnwidth]{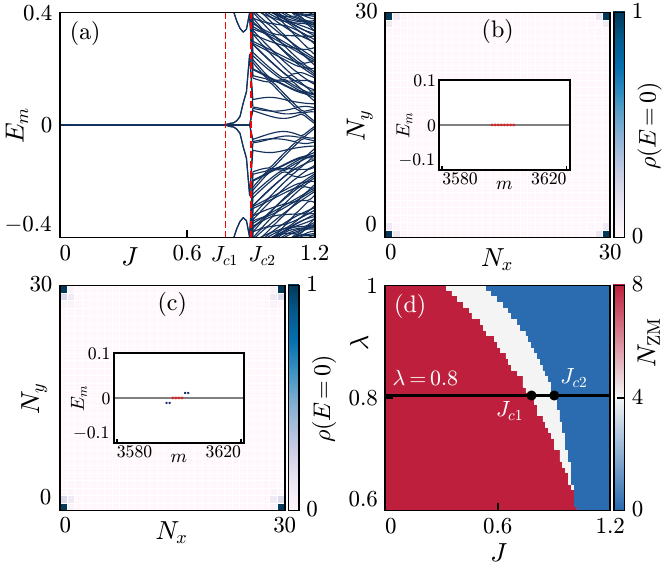}
		\caption{For an isolated meron, (a) BdG eigenenergy spectrum as a function of $J$, showing two critical points $J_{c1}\approx0.78$ and $J_{c2}\approx0.90$ that separate phases with eight and four MCMs. (b),(c) Spatially resolved LDOS at zero energy for $J=0.3t$ and $J=0.85t$, corresponding to phases with eight and four MCMs, respectively. Insets show enlarged spectra highlighting the zero-energy modes (states with $|E|<10^{-6}$ in red); each map is normalized to its maximum. (d) Number of zero-energy modes $N_{\mathrm{ZM}}$ in the $\lambda$--$J$ plane, revealing three distinct phases: a trivial phase with no zero modes (blue), and topological phases with four (white) and eight (red) MCMs. Panels (a)--(c) use $\lambda=0.8$, while $\lambda$ is varied along the vertical axis in panel (d), where the black line marks $\lambda=0.8$ and the dots indicate $J_{c1}$ and $J_{c2}$ shown in (a). Throughout this calculation, we set $\epsilon_0 = 1$, $t = 1$, $\rc = 2$, and $\gamma = \pi/2$, and the superconducting order parameter $\Delta_0$ is obtained self-consistently from the gap equation. All panels are computed for a $30\times30$ lattice with OBC.}
		\label{fig:Fig6}
	\end{figure}

We next investigate its quasiparticle spectrum by solving the BdG equations~\eqref{Eq_BdG} under OBC. The spectrum in Fig.~\ref{fig:Fig6}(a) exhibits two successive gap closings, at $J_{c1}\approx0.78$ and $J_{c2}\approx0.90$ for $\lambda=0.8$ and $\epsilon_0=t$, only quantitatively shifted from the meron--antimeron values, indicating transitions between distinct SOTSC phases and a trivial gapped superconducting phase. 
To determine the spatial character of these zero-energy states, we calculate the site-resolved LDOS using Eq.~\eqref{Eq_LDOS}. The zero-energy LDOS profiles in Figs.~\ref{fig:Fig6}(b) and \ref{fig:Fig6}(c), at $J=0.3t$ and $J=0.85t$, in the eight- and four-MCM phases, respectively, together with the corresponding eigenvalue spectra in the insets, confirm the presence of corner-localized MCMs.
To identify these phases across parameter space, we count the quasiparticle states satisfying $|E|<10^{-6}$ and plot the resulting number of zero-energy modes, $N_{\mathrm{ZM}}$, in the $\lambda$--$J$ plane, as shown in Fig.~\ref{fig:Fig6}(d). 
The phase diagram contains regions with eight and four MCMs, as well as a trivial region without zero-energy modes. 
The parameter regime supporting the SOTSC phases is comparable to that of the meron--antimeron pair.

Finally, the topological characterization by the quadrupolar winding number $N_{xy}$ in the $\rc\to0$ limit is similar to the meron--antimeron case of Fig.~\ref{fig:Fig4}, and is therefore not shown. 
Taken together, these results show that an isolated meron can support the same sequence of SOTSC and trivial phases as the meron--antimeron pair configuration. 
The pair arrangement is therefore not essential for the emergence of corner-localized MCMs.

	\section{Summary and Conclusion}\label{Sec:V}
	%======================================================
In this work, we propose a 2D MSH platform in which a quantum spin Hall insulator/$d$-wave superconductor heterostructure is coupled to a magnetic layer hosting noncollinear meron–antimeron textures, enabling tunable second-order topological superconductivity.
The QSHI induces SOC, while the magnetic layer generates a spatially modulated local exchange field in the QSHI layer. 
By treating the superconducting order self-consistently within an extended Hubbard model, we showed that an attractive nearest-neighbor interaction with vanishing on-site interaction supports a robust $d$-wave superconducting state over a broad parameter regime. 
In particular, the pairing amplitude remains finite in the presence of both a finite exchange coupling $J$ and SOC $\lambda$.
As a complementary case, discussed in Appendix~\ref{App_C}, we considered an attractive on-site interaction with vanishing nearest-neighbor interaction. 
This interaction favors an $s$-wave superconducting state that remains stable over a finite region of parameter space. 
The resulting system realizes an SOTSC phase with one MCM localized at each corner but, unlike the $d$-wave case, without a tunable change in the number of MCMs. 
Thus, the proposed platform can support HOTSC for both $d$-wave and $s$-wave pairing channels, depending on the dominant local interaction.

Focusing on the self-consistent $d$-wave state, we investigated the quasiparticle spectrum and the spatial structure of the zero-energy states in finite systems with open boundaries. 
By tuning the exchange coupling $J$, we identified three distinct regimes: an SOTSC phase hosting two MCMs at each corner, corresponding to eight MCMs in total; an SOTSC phase hosting one MCM at each corner, corresponding to four MCMs in total; and a topologically trivial fully gapped superconducting phase without MCMs. 
The transitions between these regimes are accompanied by bulk or edge-gap closings and are reflected in the evolution of the zero-energy-mode count. 
The distinct HOTSC phases are also characterized by the quadrupolar winding number $N_{xy}$ in the symmetry-preserving limit in which the corresponding chiral symmetry is well defined. 
For finite meron-core size, the out-of-plane component of the magnetic texture generally breaks this chiral symmetry; in that case, the topological phases are identified directly from the quasiparticle spectrum, the number of zero-energy modes, and their corner-resolved local density of states. 
We further showed that an isolated meron texture produces qualitatively similar SOTSC phases, indicating that the higher-order topology does not rely exclusively on the presence of a meron--antimeron pair lattice. 
To clarify the microscopic origin of the tunable transitions, we developed an effective low-energy edge theory in Appendix~\ref{App_D}. 
This theory describes the competition between the exchange-induced and superconducting edge masses and explains the change in the number of MCMs across the different topological regimes.

From an experimental perspective, meron--antimeron pairs and related bimeron textures have been realized in a variety of chiral and antiferromagnetic magnetic systems~\cite{Yu2018,Jani2021,Bhukta2024}. 
Meron pairs have also been created in in-plane magnetized films~\cite{Gao2019}, predicted in two-dimensional magnets such as CrCl$_3$~\cite{Augustin2021}, and proposed to be generated optically in antiferromagnets~\cite{Ghosh2023}. 
In parallel, magnetic layers coupled to superconducting substrates have emerged as promising platforms for engineering topological superconductivity and Majorana zero modes~\cite{Schneider2022,Chatzopoulos2021}. 
These developments suggest that heterostructures combining high-temperature $d$-wave superconductors, QSHI materials, and magnetic layers capable of supporting meronic spin textures may provide a realistic route toward realizing 2D SOTSC phases with tunable and multiple MCMs. 
More broadly, our results show that spatially structured magnetic exchange fields can control the number of MCMs in unconventional superconducting heterostructures.

%	Using the resulting self-consistent $d$-wave superconducting state, we investigated the topological properties of the finite system and identified three distinct phases characterized by different numbers of MCMs : a SOTSC phase hosting two MCMs per corner, a SOTSC phase with one MCM per corner, and a topologically trivial gapped superconducting phase without MCMs. The different higher-order topological phases are further characterized by the quadrupolar winding number $N_{xy}$. We also showed that isolated meron and antimeron textures support qualitatively similar SOTSC phase. To elucidate the origin of the tunable transition, we developed an effective low-energy edge theory in Appendix~\ref{App_D}, which captures the competition between the exchange-induced and superconducting edge masses and accounts for the change in the number of MCMs.
	
%	From the experimental perspective, meron--antimeron pair textures and related bimeron configurations have already been experimentally realized in several chiral and antiferromagnetic magnetic systems~\cite{Yu2018,Jani2021,Bhukta2024}. In addition, magnetic layers deposited on superconducting substrates have been proposed as promising platforms for realizing topological superconducting phases and Majorana zero modes~\cite{Schneider2022,Chatzopoulos2021}. Motivated by these developments, our work suggests that hybrid structures involving Fe-based superconductors, QSHI materials, and magnetic layers supporting meronic textures may provide a possible route toward realizing SOTSC phases with multiple MCMs.
	
	\subsection*{Acknowledgments}
	%======================================================
	O.A. and A.S. acknowledge SAMKHYA: High-Performance Computing Facility provided by Institute of Physics, Bhubaneswar and the two Workstations provided by Institute of Physics, Bhubaneswar from DAE APEX Project, for numerical computations. O.A., B.M., and A.K.N. acknowledge the computational resources, Kalinga cluster, at the National Institute of Science Education and Research (NISER), Bhubaneswar, India. O.A. acknowledges Udit Khanna, Amartya Pal, and Minakshi Subhadarshini for stimulating discussions.
	
	%======================================================
	
	\subsection*{Data Availability Statement} The datasets generated and analyzed during the current study are available from the corresponding author upon reasonable request.

	\appendix

	\section{Topological charge and winding properties of the magnetic textures}\label{App_A}
	
	\begin{figure*}[t]
		\centering
		\includegraphics[width=0.8\textwidth]{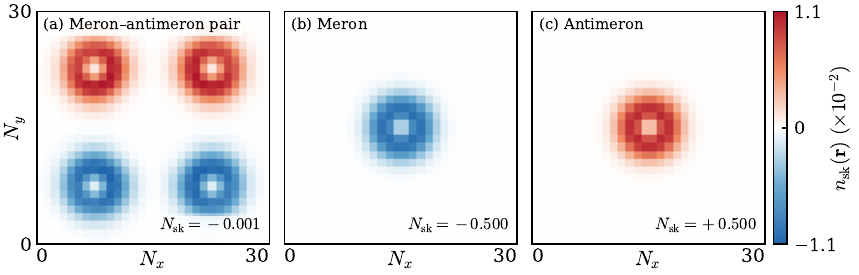}
		\caption{Topological charge density $n_{\mathrm{sk}}(\mathbf{r})$ computed 
			using the Berg--L\"{u}scher method~\cite{Berg1981} for (a) a meron--antimeron 
			pair configuration with texture centers located at the four quadrant positions 
			$\mathbf{R}_i$ (merons at $\mathbf{R}_1$ and $\mathbf{R}_2$, antimerons at $\mathbf{R}_3$ and $\mathbf{R}_4$), (b) an isolated meron with $(q, p) = (-1, -1)$, and 
			(c) an isolated antimeron with $(q, p) = (-1, 1)$. The color scale 
			represents $n_{\mathrm{sk}}(\mathbf{r})$ in units of $10^{-2}$, with red 
			(blue) corresponding to positive (negative) topological charge density. 
			The numerically obtained topological charges are 
			$\Nsk = -0.001 \approx 0$ for the meron--antimeron pair, 
			$\Nsk = -0.500$ for the meron, and 
			$\Nsk = +0.500$ for the antimeron. All panels use $r_c=3$.}
		\label{fig:AppA}
	\end{figure*}
	The spin textures considered in this work are described by the classical 
	unit vector field $\mathbf{n}(\mathbf{r}) = \mathbf{S}(\mathbf{r})/S$, 
	where the spin field $\mathbf{S}(\mathbf{r})$ is defined in 
	Eq.~(\ref{Eq_Spin_Field}), with the azimuthal and polar angle profiles 
	given by Eqs.~(\ref{Eq_Azim_Isolated}) and (\ref{Eq_Polar_Isolated}) for an isolated texture and by 
	Eqs.~(\ref{Eq_Azim_MAM}) and (\ref{Eq_Polar_MAM}) for the pair lattice. 

	The topological charge of each texture is quantified by the skyrmion 
	number~\cite{Nagaosa2013,Koshibae2016,Muhlbauer_2009,Binz_2008,Yu2018}
	\begin{equation}
		\Nsk = \int d^2\mathbf{r}\, n_{\mathrm{sk}}(\mathbf{r}),
	\end{equation}
	where the topological charge density is
	\begin{equation}
		n_{\mathrm{sk}}(\mathbf{r}) = \frac{1}{4\pi}\,\mathbf{n}(\mathbf{r})\cdot
		\bigl(\partial_x\mathbf{n}(\mathbf{r})\times
		\partial_y\mathbf{n}(\mathbf{r})\bigr).
	\end{equation}
	To see why an isolated meron or antimeron carries a fractional 
	topological charge $|\Nsk| = 1/2$, we examine the boundary conditions imposed on the 
	spin field. For an isolated texture centered at $\mathbf{R}$, the polar 
	angle profile satisfies

	\begin{equation}
		f(\mathbf{R}) = \frac{\pi}{2}(1 + p) = 
		\begin{cases} \pi & (p = +1),\\ 0 & (p = -1), \end{cases}
	\end{equation}
	and $f(\mathbf{r}) \to \pi/2$ as $|\mathbf{r} - \mathbf{R}| \to \infty$. 
	Consequently, the out-of-plane spin component satisfies $S_z = \cos f = -p$ 
	at the core and $S_z = 0$ at infinity, meaning the spin field covers only 
	a hemisphere of the unit sphere rather than the full sphere. Since the 
	skyrmion number counts the number of times the unit vector field wraps 
	around the full sphere, a texture covering only one hemisphere contributes 
	$|\Nsk| = 1/2$, in contrast to a full skyrmion for which 
	$f$ varies from $0$ to $\pi$ and $\Nsk = \pm 1$. The sign of 
	$\Nsk$ is determined by the combined effect of the azimuthal 
	winding $q$ and the core orientation $p$, as summarized in 
	Table~\ref{tab:textures} in Sec.~\ref{Sec:II} of the main text: for the profiles of Eqs.~\eqref{Eq_Azim_Isolated} and~\eqref{Eq_Polar_Isolated}, the integral gives $\Nsk=-qp/2$.
	We verify these topological charges numerically on a $30\times 30$ 
	lattice using the Berg--L\"{u}scher 
	approach~\cite{Berg1981}, which provides a topologically protected 
	discretization. Each lattice plaquette is divided into two triangles, oriented counter-clockwise in the $(x,y)$ plane, 
	and the contribution of each triangle to $\Nsk$ is $\Omega/4\pi$, where $\Omega$ is 
	the solid angle subtended by the three spin vectors at its 
	corners on the unit sphere,
	\begin{equation}
		\tan\frac{\Omega}{2} = 
		\frac{\mathbf{n}_1\cdot(\mathbf{n}_2\times\mathbf{n}_3)}
		{1 + \mathbf{n}_1\cdot\mathbf{n}_2 + 
			\mathbf{n}_2\cdot\mathbf{n}_3 + 
			\mathbf{n}_1\cdot\mathbf{n}_3},
	\end{equation}
	where $\mathbf{n}_1$, $\mathbf{n}_2$, $\mathbf{n}_3$ are the unit spin 
	vectors at the three corners. The total topological charge, 
	$\Nsk=\frac{1}{4\pi}\sum\Omega$, is obtained by summing over all triangles. This method yields well-quantized 
	values already on coarse lattices, making it more reliable than 
	the finite-difference approach for discrete spin textures.
	
	The spatial profiles of the topological charge density $n_{\mathrm{sk}}(\mathbf{r})$ for the meron--antimeron pair, isolated meron, and isolated antimeron configurations are shown in Fig.~\ref{fig:AppA}. The numerically obtained values $\Nsk \approx 0$, $-1/2$, and $+1/2$, respectively, are in excellent agreement with the analytically expected quantized values, confirming the topological nature of the spin textures considered in this work. For the pair lattice, the two merons ($-1/2$ each) and the two antimerons ($+1/2$ each) cancel, so that $\Nsk=0$. The small deviations from exact quantization arise from the finite lattice and the open boundary, where $f$ reaches $\pi/2$ only asymptotically, so that $S_z$ does not vanish exactly at the edge.

	\vspace{.3pt}

\section{Bogoliubov--de Gennes transformation and self-consistent determination of the order parameter}\label{App_SC}

In this appendix, following Ref.~\cite{Zhu2016BdG}, we use the Bogoliubov transformation to derive the BdG equations~\eqref{Eq_BdG} and the self-consistency condition~\eqref{Eq_GapEq}, and we describe the numerical procedure used to solve them self-consistently.

For $U=0$, the interaction in Eq.~\eqref{Eq_Hint} of the main text reads
\begin{equation}
    H_{\mathrm{int}}
    =
    -\frac{V}{2}
    \sum_{\langle \mathbf r_i,\mathbf r_j\rangle,\alpha,\sigma,\sigma'}
    n_{\mathbf r_i\alpha\sigma}\,n_{\mathbf r_j\alpha\sigma'} .
\end{equation}
Within mean-field theory, this interaction leads to the Hamiltonian $H_{\mathrm{MF}}$ of Eq.~\eqref{Eq_MF}, in which the pairing term is summed over the NN pairs $(\mathbf r_i,\mathbf r_j)$ in both orientations, with $\Delta(\mathbf r_i,\mathbf r_j)=\Delta(\mathbf r_j,\mathbf r_i)$. To obtain the BdG equations, we use the equation of motion for the electron operators. We collect the operators of a site into $c_{\mathbf r_i}=(c_{a\uparrow},c_{a\downarrow},c_{b\uparrow},c_{b\downarrow})^T_{\mathbf r_i}$ and $\bar c_{\mathbf r_i}=(c^{\dagger}_{a\uparrow},c^{\dagger}_{a\downarrow},c^{\dagger}_{b\uparrow},c^{\dagger}_{b\downarrow})^T_{\mathbf r_i}$, so that $\Psi_{\mathbf r_i}=(c_{\mathbf r_i},\bar c_{\mathbf r_i})^T$. Using $[A,BC]=\{A,B\}C-B\{A,C\}$ and the anticommutation relations of $c$ and $c^{\dagger}$, the commutator of an annihilation operator with the hopping part of $H_{\mathrm{MF}}$ gives $\sum_{\mathbf r_j}h_{\mathbf r_i,\mathbf r_j}c_{\mathbf r_j}$. In the pairing term, only the terms containing the creation operator at $\mathbf r_i$ contribute, for example
\begin{equation}
\begin{aligned}
    \Bigl[c_{\mathbf r_i\alpha\uparrow},\,\Delta(\mathbf r_i,\mathbf r_j)\,c^{\dagger}_{\mathbf r_i\alpha\uparrow}c^{\dagger}_{\mathbf r_j\alpha\downarrow}\Bigr]
    &=\Delta(\mathbf r_i,\mathbf r_j)\,c^{\dagger}_{\mathbf r_j\alpha\downarrow},
    \\
    \Bigl[c_{\mathbf r_i\alpha\downarrow},\,\Delta(\mathbf r_j,\mathbf r_i)\,c^{\dagger}_{\mathbf r_j\alpha\uparrow}c^{\dagger}_{\mathbf r_i\alpha\downarrow}\Bigr]
    &=-\Delta(\mathbf r_i,\mathbf r_j)\,c^{\dagger}_{\mathbf r_j\alpha\uparrow},
\end{aligned}
\end{equation}
while the Hermitian-conjugate part of the pairing term contains only annihilation operators and commutes with $c_{\mathbf r_i\alpha\sigma}$. Collecting all terms in the Nambu spinor, we obtain
\begin{equation}
    [\Psi_{\mathbf r_i},H_{\mathrm{MF}}]=\sum_{\mathbf r_j}\mathcal H_{\mathrm{BdG}}(\mathbf r_i,\mathbf r_j)\,\Psi_{\mathbf r_j},
    \label{Eq_comm_app}
\end{equation}
where $\mathcal H_{\mathrm{BdG}}$ is the matrix of Eq.~\eqref{Eq_BdG_ham}.

We now introduce the Bogoliubov transformation, which expresses the electron operators in terms of quasiparticle operators $\gamma_n$,
\begin{equation}
    c_{\mathbf r_i\alpha\sigma}
    =
    \sum_{n}{}^{'}
    \left[
    u^{n}_{\alpha\sigma}(\mathbf r_i)\,\gamma_n
    +
    v^{n*}_{\alpha\sigma}(\mathbf r_i)\,\gamma_n^{\dagger}
    \right],
\end{equation}
where the prime indicates a sum over the states with $E_n>0$, and the fermionic operators satisfy $\{\gamma_n,\gamma_{m}^{\dagger}\}=\delta_{nm}$ and $\{\gamma_n,\gamma_{m}\}=\{\gamma_n^{\dagger},\gamma_{m}^{\dagger}\}=0$. The amplitudes $u^n_{\alpha\sigma}(\mathbf r_i)$ and $v^n_{\alpha\sigma}(\mathbf r_i)$, and the energies $E_n$, are to be determined such that $H_{\mathrm{MF}}$ takes the diagonal form
\begin{equation}
    H_{\mathrm{MF}}=E_g+\sum_{n}{}^{'}E_n\,\gamma_n^{\dagger}\gamma_n ,
    \label{Eq_diag_app}
\end{equation}
where $E_g$ is a constant. In the Nambu basis the transformation reads $\Psi_{\mathbf r_i}=\sum_n^{'}\left[\Phi^n(\mathbf r_i)\gamma_n+\tilde\Phi^n(\mathbf r_i)\gamma_n^{\dagger}\right]$, with $\Phi^n=(u^n,v^n)^T$ and $\tilde\Phi^n=(v^{n*},u^{n*})^T$.

Equation~\eqref{Eq_diag_app} implies $[\gamma_n,H_{\mathrm{MF}}]=E_n\gamma_n$ and $[\gamma_n^{\dagger},H_{\mathrm{MF}}]=-E_n\gamma_n^{\dagger}$. The commutator of $\Psi_{\mathbf r_i}$ with $H_{\mathrm{MF}}$ can then also be written using the transformation,
\begin{equation}
    [\Psi_{\mathbf r_i},H_{\mathrm{MF}}]
    =
    \sum_n{}^{'}E_n\left[\Phi^n(\mathbf r_i)\gamma_n-\tilde\Phi^n(\mathbf r_i)\gamma_n^{\dagger}\right],
\end{equation}
while Eq.~\eqref{Eq_comm_app} gives
\begin{equation}
\begin{aligned}
    [\Psi_{\mathbf r_i},H_{\mathrm{MF}}]
    =
    \sum_n{}^{'}\sum_{\mathbf r_j}&\mathcal H_{\mathrm{BdG}}(\mathbf r_i,\mathbf r_j)
    \\
    &\times\left[\Phi^n(\mathbf r_j)\gamma_n+\tilde\Phi^n(\mathbf r_j)\gamma_n^{\dagger}\right].
\end{aligned}
\end{equation}
Since the operators $\gamma_n$ and $\gamma_n^{\dagger}$ are independent, we equate their coefficients. Those of $\gamma_n$ give
\begin{equation}
    \sum_{\mathbf r_j}\mathcal H_{\mathrm{BdG}}(\mathbf r_i,\mathbf r_j)\,\Phi^{n}(\mathbf r_j)=E_n\,\Phi^{n}(\mathbf r_i),
\end{equation}
which is the BdG equation~\eqref{Eq_BdG}, while those of $\gamma_n^{\dagger}$ give $\sum_{\mathbf r_j}\mathcal H_{\mathrm{BdG}}(\mathbf r_i,\mathbf r_j)\tilde\Phi^{n}(\mathbf r_j)=-E_n\tilde\Phi^{n}(\mathbf r_i)$. The latter follows from the former because the singlet pairing block $\Delta(\mathbf r_i,\mathbf r_j)(\sigma_0\otimes is_y)$ is antisymmetric under the exchange of its two particles, so that $\mathcal H_{\mathrm{BdG}}$ obeys the particle--hole symmetry $\Sigma\mathcal H_{\mathrm{BdG}}^{*}\Sigma=-\mathcal H_{\mathrm{BdG}}$, with $\Sigma=\tau_x\sigma_0 s_0$, and $\tilde\Phi^n=\Sigma\Phi^{n*}$. The $8N_xN_y$ eigenstates of $\mathcal H_{\mathrm{BdG}}$ therefore come in pairs $(E_{\bar n},u^{\bar n},v^{\bar n})=(-E_n,v^{n*},u^{n*})$: the states with $E_n>0$ define the quasiparticles, while those with $E_n<0$ are their partners and contain no additional information.

Next, we compute the pair correlation function that enters the order parameter. As in the main text, we suppress the orbital index $\alpha$ of the amplitudes $u$ and $v$, since $\Delta$ is the same in both orbital sectors. Since $H_{\mathrm{MF}}$ is diagonal in terms of the $\gamma_n$, the quasiparticle occupations are $\langle\gamma_n^{\dagger}\gamma_m\rangle=\delta_{nm}f(E_n)$ and $\langle\gamma_n\gamma_m\rangle=0$, where $f$ is the Fermi function. Inserting the Bogoliubov transformation into the pair correlation function gives
\begin{equation}
\begin{aligned}
    \langle c_{\mathbf r_i\alpha\uparrow}c_{\mathbf r_j\alpha\downarrow}\rangle
    =
    \sum_n{}^{'}
    \Bigl\{&
    u^n_{\uparrow}(\mathbf r_i)\,v^{n*}_{\downarrow}(\mathbf r_j)\,\bigl[1-f(E_n)\bigr]
    \\
    &+
    v^{n*}_{\uparrow}(\mathbf r_i)\,u^{n}_{\downarrow}(\mathbf r_j)\,f(E_n)
    \Bigr\}.
\end{aligned}
\end{equation}
The second term can be rewritten in terms of the partner state $\bar n$. Using $v^{n*}_{\uparrow}=u^{\bar n}_{\uparrow}$, $u^{n}_{\downarrow}=v^{\bar n*}_{\downarrow}$, and $f(E_n)=f(-E_{\bar n})=1-f(E_{\bar n})$, we find
\begin{equation}
    v^{n*}_{\uparrow}(\mathbf r_i)\,u^{n}_{\downarrow}(\mathbf r_j)\,f(E_n)
    =
    u^{\bar n}_{\uparrow}(\mathbf r_i)\,v^{\bar n*}_{\downarrow}(\mathbf r_j)\,\bigl[1-f(E_{\bar n})\bigr],
\end{equation}
which is the first term evaluated for the partner state, with $E_{\bar n}<0$. Hence the first term summed over the states with $E_n>0$, plus the same expression summed over their partners with $E_{\bar n}<0$, is this single expression summed over all eigenstates. The prime is therefore dropped, and from here on $n$ runs over all $8N_xN_y$ eigenstates,
\begin{equation}
\begin{aligned}
    \langle c_{\mathbf r_i\alpha\uparrow}c_{\mathbf r_j\alpha\downarrow}\rangle
    &=
    \sum_n u^n_{\uparrow}(\mathbf r_i)\,v^{n*}_{\downarrow}(\mathbf r_j)\,\bigl[1-f(E_n)\bigr]
    \\
    &=
    \frac{1}{2}\sum_n u^n_{\uparrow}(\mathbf r_i)\,v^{n*}_{\downarrow}(\mathbf r_j)\tanh\!\left(\frac{E_n}{2T}\right),
\end{aligned}
\end{equation}
where we used $1-f(E_n)=\tfrac12[1+\tanh(E_n/2T)]$ and the completeness of the eigenvectors, $\sum_n u^n_{\sigma}(\mathbf r_i)v^{n*}_{\sigma'}(\mathbf r_j)=0$, which removes the constant term. The order parameter is $\Delta(\mathbf r_i,\mathbf r_j)=\tfrac{V}{2}\left[\langle c_{\mathbf r_i\alpha\uparrow}c_{\mathbf r_j\alpha\downarrow}\rangle-\langle c_{\mathbf r_i\alpha\downarrow}c_{\mathbf r_j\alpha\uparrow}\rangle\right]$. Using the anticommutation of the fermion operators, $\langle c_{\mathbf r_i\alpha\downarrow}c_{\mathbf r_j\alpha\uparrow}\rangle=-\langle c_{\mathbf r_j\alpha\uparrow}c_{\mathbf r_i\alpha\downarrow}\rangle$, and the result above for both correlators, we obtain
\begin{equation}
\begin{aligned}
    \Delta(\mathbf r_i,\mathbf r_j)
    &=
    \frac{V}{2}\Bigl[\langle c_{\mathbf r_i\alpha\uparrow}c_{\mathbf r_j\alpha\downarrow}\rangle+\langle c_{\mathbf r_j\alpha\uparrow}c_{\mathbf r_i\alpha\downarrow}\rangle\Bigr]
    \\
    &=
    \frac{V}{4}\sum_n
    \Bigl[
    u^n_{\uparrow}(\mathbf r_i)\,v^{n*}_{\downarrow}(\mathbf r_j)
    \\
    &\hspace{3.5em}
    +
    u^n_{\uparrow}(\mathbf r_j)\,v^{n*}_{\downarrow}(\mathbf r_i)
    \Bigr]
    \tanh\!\left(\frac{E_n}{2T}\right),
\end{aligned}
\end{equation}
which is the self-consistency condition of Eq.~\eqref{Eq_GapEq}.

The self-consistent solution is obtained numerically as follows. For an $N_x\times N_y$ lattice, the BdG matrix has dimension $8N_xN_y$ ($7200$ for $30\times30$), and, under OBC, the unknowns are the $N_x(N_y-1)+N_y(N_x-1)=2N_xN_y-N_x-N_y$ bond amplitudes $\Delta(\mathbf r_i,\mathbf r_i+\hat x)$ and $\Delta(\mathbf r_i,\mathbf r_i+\hat y)$. The bonds across the boundary, which would bring this number to $2N_xN_y$, are absent. The normal-state Hamiltonian, including the spin texture, does not change during the iteration and is built once.

	\begin{figure*}[t]
		\centering
		\includegraphics[width=0.7\textwidth]{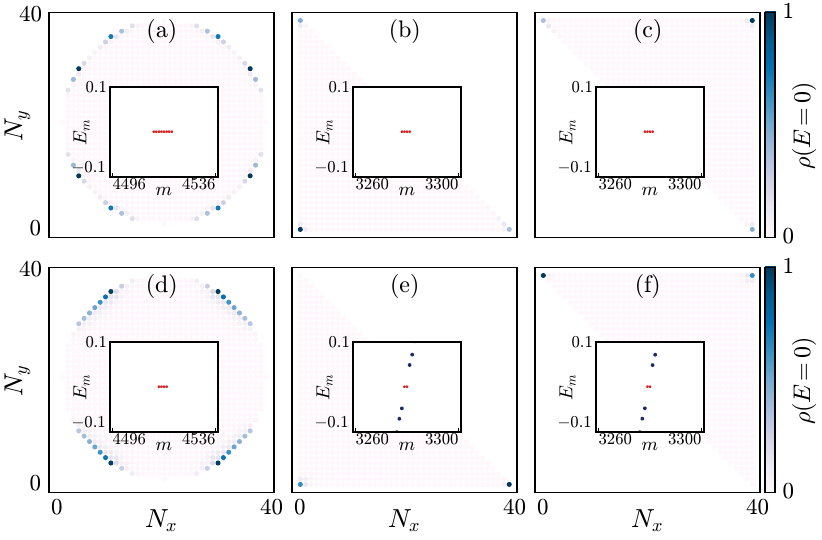}
		
		\caption{Robustness of MCMs against geometric deformations of the 
			boundary for the meron--antimeron pair configuration. The color scale represents the normalized zero-energy LDOS $\rho$ and the insets in each panel show the BdG eigenspectrum $E_{m}$ as a function of state index $m$ around zero energy (states with $|E|<10^{-6}$ in red). Top row: eight-MCM regime ($J = 0.3t$) for (a) circular 
			geometry, showing 8 high-intensity LDOS spots 
			around the periphery with the inset showing 8 zero-energy 
			states, (b) lower triangular geometry with MCMs localized at three 
			corners, and (c) upper triangular geometry with MCMs localized at 
			another three corners. Bottom row: four-MCM regime ($J = 1.5t$) 
			for (d) circular geometry showing 4 high-intensity LDOS spots around 
			the periphery with the inset showing 4 zero-energy states, 
			(e) lower triangular geometry with MCMs localized at two corners, and 
			(f) upper triangular geometry with MCMs localized at the other 
			two corners. In all panels the system is cut from a $40\times 40$ lattice 
			with a uniform $d$-wave pairing amplitude $\Delta_0=0.5t$, $\epsilon_0=t$, $\lambda=t$ and $\rc=2$.}
		\label{fig:AppB}
	\end{figure*}
\begin{enumerate}
\item We start from a non-uniform set of bond amplitudes $\{\Delta^{(0)}(\mathbf r_i,\mathbf r_j)\}$. Each amplitude is drawn independently from a Gaussian random-number generator, and the set is scaled such that its largest magnitude is $0.5$. A random start does not bias the iteration toward a particular pairing symmetry. The scale $0.5$ is of the order of the converged gap amplitude, and it is large enough to avoid the trivial solution $\Delta=0$.
\item At iteration $k$, we construct $\mathcal H_{\mathrm{BdG}}$ from the set $\{\Delta^{(k)}\}$ and obtain all its eigenvalues and eigenvectors by full diagonalization with a dense Hermitian eigensolver (LAPACK, via NumPy/SciPy). We then evaluate Eq.~\eqref{Eq_GapEq} on every bond, which gives a new set $\{\Delta_{\rm out}(\mathbf r_i,\mathbf r_j)\}$.
\item After each iteration, the input for the next one is the linear mixture of the new and the old sets, $\Delta^{(k+1)}=w\,\Delta_{\rm out}+(1-w)\,\Delta^{(k)}$, with $w=0.3$. Since Eq.~\eqref{Eq_GapEq} is nonlinear in $\Delta$, directly substituting $\Delta^{(k+1)}=\Delta_{\rm out}$ can lead to oscillations or divergence of the iteration; a mixing parameter $w<1$ damps them and stabilizes the convergence.
\item Steps 2 and 3 are repeated until the largest difference between the two sets, $\varepsilon=\max_{\rm bonds}\left|\Delta_{\rm out}-\Delta^{(k)}\right|$, falls below $10^{-6}$.
\item From the converged bond amplitudes, we compute the local $d$-wave amplitude $\Delta_0(\mathbf r_i)$ and its spatial average $\bar\Delta_0$ [Eq.~\eqref{Eq_avg_order_param}].
\end{enumerate}
% OPTIONAL (only if Anderson acceleration is used for the reported data), add to step 3(e):
% "For faster convergence, the linear mixing is replaced after five iterations by Anderson acceleration with a history of five steps."

\section{Majorana Corner Modes in Disc and Triangular Geometries}\label{App_B}
	
	To assess the robustness of the SOTSC phase against geometric 
	deformations, we examine the spatial distribution of the zero-energy 
	LDOS and the eigenspectrum of the BdG Hamiltonian~\eqref{Eq_BdG_ham} under two alternative boundary geometries, a circular and a triangular truncation of the square lattice. In both cases, the bulk Hamiltonian is that of the main text, with a uniform $d$-wave pairing amplitude $\Delta(\mathbf r,\mathbf r+\hat x)=-\Delta(\mathbf r,\mathbf r+\hat y)=\Delta_0=0.5t$, $\epsilon_0=t$, $\lambda=t$ and $\rc=2$, on a $40\times40$ lattice; only 
	the boundary shape is modified by removing the lattice sites outside 
	the desired geometry. Since the pairing amplitude is held fixed rather than determined self-consistently, the critical couplings differ from those of Fig.~\ref{fig:Fig3}.
	Figure~\ref{fig:AppB} shows the zero-energy LDOS and the corresponding 
	eigenspectrum for both the eight-MCM regime ($J = 0.3t$) and the 
	four-MCM regime ($J = 1.5t$) under circular and triangular geometries. 
	In the circular geometry, the zero-energy LDOS exhibits 8 (4) discrete high-intensity spots distributed around the circular periphery in the eight-MCM (four-MCM) regime, with some residual weight along the intervening boundary segments. These spots correspond to the curved remnants of the original square corners, confirming that the MCMs remain predominantly localized at corner-like regions even in the absence of sharp corners. The eigenspectrum insets confirm that these modes stay at $|E|\lesssim6\times10^{-5}$ in the eight-MCM regime and $|E|\lesssim2\times10^{-5}$ in the four-MCM regime, well separated from the bulk states. For the triangular geometry, the boundary cut removes one corner of the square lattice. In the lower triangular geometry, the upper-right corner is removed, and the surviving MCMs localize at the remaining three corners. In the upper triangular geometry, the lower-left corner is removed, and the surviving MCMs localize at the another set of three corners. In the eight-MCM regime, four MCMs survive under triangular masking, while in the four-MCM regime, two MCMs survive. 
	The reduction in the number of MCMs under triangular truncation — from 8 to 4 in the eight-MCM regime and from 4 to 2 in the four-MCM regime — confirms that the corner-localized zero-energy modes persist under geometric deformations of the boundary, although their number is reduced. A complete understanding of the precise reduction in the MCM count would require an edge theory analysis, which is left for future work.

\begin{figure*}[t]
	\centering
	\includegraphics[width=0.95\textwidth]{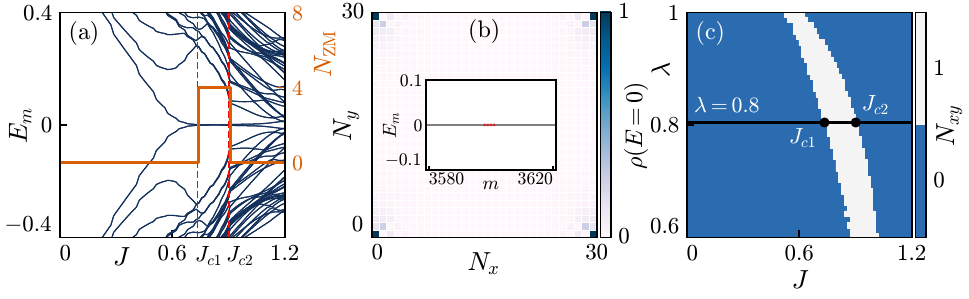}
	\caption{Topological properties of the meron--antimeron pair texture 
		in the presence of a conventional $s$-wave superconductor proximitized 
		to the QSHI layer. 
		(a) Quasiparticle spectrum $E_m$ as a function of exchange coupling 
		$J$ at $\lambda = 0.8$, computed with OBC along both $x$ and $y$ 
		directions, together with the number of zero-energy modes $N_{\mathrm{ZM}}$ (orange line, right axis). 
		The spectrum is gapped for small $J$; four states reach zero energy between $J_{c1}\approx0.74$ and $J_{c2}\approx0.90$, 
		beyond which the system becomes gapless as the pairing amplitude vanishes. 
		(b) Zero-energy LDOS $\rho$ at $J = 0.82t$, showing four MCMs 
		localized at the corners of the system with one Majorana mode per 
		corner. The inset shows the quasiparticle eigenspectrum $E_m$ as a 
		function of state index $m$, with the four zero-energy states ($|E|<10^{-6}$) in red. 
		(c) Quadrupolar winding number $N_{xy}$ in the $\lambda$--$J$ plane, 
		with $N_{xy} = 1$ (white) denoting the phase with four MCMs (one per corner) and $N_{xy} = 0$ (blue) the trivial phase; 
		the black line marks $\lambda=0.8$ and the dots indicate $J_{c1}$ and $J_{c2}$ shown in (a). 
		Throughout, $\epsilon_0 = 1$, 
		$t = 1$ and $\rc = 2$, on a $30\times30$ lattice, and $\Delta_s$ is obtained self-consistently for $U=-3t$ and $V=0$.}
	\label{fig:AppC}
\end{figure*}

	\section{Meron--antimeron texture with $s$-wave superconductivity}
	\label{App_C}
	
	In this appendix, we discuss the topological properties of the 
	meron--antimeron texture in the presence of a conventional $s$-wave 
	superconductor proximitized to the QSHI layer, in which the on-site pairing amplitude $\Delta_s(\mathbf r_i)$ is generated self-consistently by the attractive interaction $U$. The normal-state 
	Hamiltonian is $H_0$ of Eq.~\eqref{eq:H_0} in Sec.~\ref{Sec:II}, the same as in the main text. We now set $V=0$, so that only the on-site 
	interaction of Eq.~\eqref{Eq_Hint} remains,
	\begin{equation}
		H_{\mathrm{int}}
		= U \sum_{\mathbf r_i,\alpha} n_{\mathbf r_i\alpha\uparrow} n_{\mathbf r_i\alpha \downarrow},
		\qquad U<0 .
	\end{equation}
	A mean-field decoupling of this interaction in the pairing channel, using 
	$n_{\mathbf r_i\alpha\uparrow}n_{\mathbf r_i\alpha\downarrow}=c^{\dagger}_{\mathbf r_i\alpha\uparrow}c^{\dagger}_{\mathbf r_i\alpha\downarrow}c_{\mathbf r_i\alpha\downarrow}c_{\mathbf r_i\alpha\uparrow}$, gives
	\begin{equation}
		\begin{aligned}
		H_{\mathrm{MF}}^{s}
		=
		&H_0
		+
		\sum_{\mathbf r_i,\alpha}
		\left(
		\Delta_s (\mathbf r_i)\,
		c^{\dagger}_{\mathbf r_i\alpha\uparrow}
		c^{\dagger}_{\mathbf r_i\alpha\downarrow}
		+ \mathrm{H.c.}
		\right)
		\\
		&
		-
		\sum_{\mathbf r_i,\alpha}
		\frac{|\Delta_s (\mathbf r_i)|^2}{U},
		\end{aligned}
		\label{Eq_MF_s}
	\end{equation}
	where the on-site $s$-wave order parameter is
	\begin{equation}
		\Delta_s (\mathbf r_i)
		=
		U
		\langle
		c_{\mathbf r_i\alpha\downarrow}
		c_{\mathbf r_i\alpha\uparrow}
		\rangle .
		\label{Eq_Delta_s}
	\end{equation}
	As in the main text, the order parameter is assumed to be identical in both orbital sectors, so that $\Delta_s(\mathbf r_i)$ does not carry an orbital index. With the Nambu spinor $\Psi_{\mathbf r_i}$ of Sec.~\ref{Sec:II}, the BdG Hamiltonian takes the form of Eq.~\eqref{Eq_BdG_ham} with an on-site pairing term,
	\begin{equation}
		\mathcal H_{\mathrm{BdG}}^{s}
		=
		\begin{pmatrix}
			h
			&
			\Delta_s (\mathbf r_i)\,(\sigma_0 \, is_y)
			\\[4pt]
			\Delta_s (\mathbf r_i)^{*}\,(\sigma_0 \, is_y)^{\dagger}
			&
			-h^{*}
		\end{pmatrix},
		\label{eq:H_BdG_s}
	\end{equation}
	where $h$ is the normal-state block of $H_0$ and the pairing block is diagonal in the site index, i.e. $\Delta(\mathbf r_i,\mathbf r_j)\to\Delta_s(\mathbf r_i)\delta_{\mathbf r_i,\mathbf r_j}$. 
	We consider the same meron--antimeron pair texture as in the main 
	text, with texture centers located at the four quadrant positions 
	$\mathbf{R}_i$ and winding numbers $(q_1,q_2,q_3,q_4) = 
	(+1,-1,-1,+1)$ and core orientation parameters 
	$(p_1,p_2,p_3,p_4) = (+1,-1,+1,-1)$, as defined in 
	Sec.~\ref{Sec:II}. 
	
	The Bogoliubov transformation of Appendix~\ref{App_SC} leads, with the pairing block $\Delta_s(\mathbf r_i)\,\delta_{\mathbf r_i,\mathbf r_j}\,(\sigma_0\, is_y)$ in Eq.~\eqref{Eq_BdG_ham}, to the same BdG equations, 
	$\sum_{\mathbf r_j}\mathcal H_{\mathrm{BdG}}(\mathbf r_i,\mathbf r_j)\Phi^n(\mathbf r_j)=E_n\Phi^n(\mathbf r_j)$. The on-site pair correlation function follows from the quasiparticle occupations exactly as for the bond correlator, 
	$\langle c_{\mathbf r_i\downarrow}c_{\mathbf r_i\uparrow}\rangle=-\tfrac12\sum_n u^n_{\uparrow}(\mathbf r_i)\,v^{n*}_{\downarrow}(\mathbf r_i)\tanh(E_n/2T)$, 
	where the orbital index is dropped as in Sec.~\ref{Sec:II} (the components refer to a single orbital sector) and $n$ runs over all eigenstates. Eq.~\eqref{Eq_Delta_s} then gives the gap equation
	\begin{equation}
		\Delta_s (\mathbf r_i)
		=
		-\frac{U}{2}
		\sum_n
		u^n_{\uparrow}(\mathbf r_i)\,v^{n*}_{\downarrow}(\mathbf r_i)
		\tanh\!\left(\frac{E_n}{2T}\right).
		\label{Eq_GapEq_s}
	\end{equation}
	It is solved self-consistently with the iterative procedure of Appendix~\ref{App_SC}, in which the unknowns are now the $N_xN_y$ on-site amplitudes $\Delta_s(\mathbf r_i)$ instead of the bond amplitudes. We take $U=-3t$ and $V=0$, at $T=0$. 
	The converged solution gives the spatially varying order parameter $\Delta_s(\mathbf r_i)$ in the presence of the meron--antimeron texture. With it, we diagonalize the BdG Hamiltonian~\eqref{eq:H_BdG_s} for a $30\times30$ lattice with OBC, using $\epsilon_0=1$, $t=1$, and $\rc=2$, and, as in the main text, we analyze the quasiparticle spectrum, the zero-energy LDOS, and the number of zero-energy modes $N_{\mathrm{ZM}}$, as well as the topological invariant $N_{xy}$ in the limit $\rc\to0$. Here $N_{\mathrm{ZM}}$ counts the states with $|E|<10^{-6}$.
	
	Figure~\ref{fig:AppC}(a) shows the quasiparticle spectrum $E_m$ 
	as a function of exchange coupling $J$ at $\lambda=0.8$, computed with OBC along both $x$ and $y$ directions. 
	The spectrum is gapped for small $J$ and the gap decreases steadily with $J$. Near $J_{c1}\approx0.74$ four states merge to zero energy 
	and stay there up to $J_{c2}\approx0.90$, where the pairing amplitude vanishes and the system becomes gapless; the number of zero-energy modes $N_{\mathrm{ZM}}$ is 
	shown by the orange line. The zero-energy LDOS 
	$\rho$ computed at a representative value of $J=0.82t$ within this 
	topological regime is shown in Fig.~\ref{fig:AppC}(b), confirming 
	that four MCMs are localized at the corners of the system, with one 
	Majorana mode per corner. The inset shows the corresponding 
	eigenspectrum, with four zero-energy states ($|E|<10^{-6}$) separated from the bulk states by a gap of about $0.2$.
	To map out the phase boundaries, we determine the quadrupolar winding number $N_{xy}$ across the $\lambda$--$J$ plane, where $N_{xy}=1$ corresponds to one MCM per corner. 
	The resulting phase diagram, with isotropic SOC 
	$\lambda_x = \lambda_y \equiv \lambda$, is shown in 
	Fig.~\ref{fig:AppC}(c). The white band corresponds to the 
	topological phase with $N_{xy} = 1$ (four MCMs), while the blue region 
	corresponds to the trivial phase with $N_{xy} = 0$. The band shifts to lower $J$ with increasing $\lambda$, and no region with $N_{xy}=2$ appears.

	Taken together, these results establish that the $s$-wave 
	proximitized system hosts four MCMs in a single window of exchange coupling, without a tunable 
	transition in their number. This stands in contrast to the 
	self-consistent $d$-wave case studied in the main text, where a 
	parameter-tunable transition between eight and four MCMs is 
	observed. The absence of this tunability in the $s$-wave case 
	indicates that the tunability found in the main text 
	is linked to the $d$-wave pairing symmetry rather 
	than to the magnetic texture alone.
	
	\section{Effective low-energy model and edge theory}
	\label{App_D}
	
	In this appendix, following the approach of Refs.~\cite{Minakshi2025,Pritam2_2024,Das2026}, 
	generalized to spatially varying textures and self-consistently determined 
	pairing, we derive an effective low-energy model for our system in the 
	$\rc \to 0$ limit and construct the corresponding edge theory and MCM solutions. 
	In this limit, the out-of-plane spin component vanishes ($S_z=0$) and the 
	texture reduces to a purely in-plane configuration with local orientation 
	$\theta(\mathbf{r}) = \phi(\mathbf{r}) + \gamma$, where $\phi(\mathbf{r})$ 
	is given by Eq.~(\ref{Eq_Azim_MAM}).
	
	\subsection{Local spin rotation and effective Hamiltonian}
	\label{App_D1}
	
	We start from the low-energy continuum Hamiltonian corresponding to 
	Eq.~(\ref{Eq_BdG_ham}), written in the mixed real- and momentum-space 
	representation,
	\begin{equation}
		\begin{aligned}
		H_{L} = {}&\xi_{k}\,\tau_z\Gamma_1 + \Delta_d\,\tau_y\Gamma_3 
		+ 2\lambda_x k_x\,\tau_0\Gamma_5 + 2\lambda_y k_y\,\tau_z\Gamma_6
		\\
		&+ J\,(S_x\, \tau_z\Gamma_2 + S_y\, \tau_0\Gamma_3) ,
		\end{aligned}
		\label{Eq_HL}
	\end{equation}
	where $\xi_k = \epsilon_0 - 4t + t(k_x^2 + k_y^2)$, 
	$\Delta_d = \bar{\Delta}_0 (k_x^2 - k_y^2)$ is the kinetic term and $d$-wave pairing with 
	the amplitude set by the converged self-consistent order parameter 
	$\bar{\Delta}_0$ [Eq.~(\ref{Eq_avg_order_param})], respectively. Here $\tau_i$, $\sigma_i$, $s_i$ are Pauli matrices in the particle-hole, orbital, and spin spaces, and $\Gamma_1,\dots,\Gamma_6$ are the matrices of Sec.~\ref{Sec:II}; the products $\tau_i\Gamma_j$ act in the conventional Nambu basis of Sec.~\ref{Sec:II}.
	The exchange term in Eq.~(\ref{Eq_HL}) is spatially nonuniform and given in Eq.~\eqref{Eq_Spin_Field} in Sec.~\ref{Sec:II} with $S_z=0$. We align it by 
	the local unitary transformation~\cite{Hess2022}
	\begin{equation}
		R = \sigma_0\, e^{-\frac{i}{2}\theta(\mathbf{r})\, \tau_z s_z},
		\label{Eq_U_rotation}
	\end{equation}
	which rotates the local exchange field at every point onto the $+\hat{x}$ 
	spin axis. We have verified on the lattice that this transformation 
	renders the exchange term exactly uniform, 
	$R^{\dagger}\big[J(S_x\,\tau_z\Gamma_2 + S_y\,\tau_0\Gamma_3)\big]R = J\,\tau_z\Gamma_2$, at 
	every site. The effective Hamiltonian 
	$H_{\mathrm{eff}} = R^{\dagger} H_{L} R$ then takes the form
	\begin{align}
		H_{\mathrm{eff}} 
		&= \xi_{k,\mathrm{eff}}\,\tau_z\Gamma_1 + \Delta_{d,\mathrm{eff}}
		+ J\,\tau_z\Gamma_2 
		\nonumber\\
		&\quad
		+ 2\lambda_x k_x\,\tau_0\Gamma_5 + 2\lambda_y k_y\,\tau_z\Gamma_6
		- t\left(\boldsymbol{\nabla}\phi\cdot\mathbf{k}\right)\tau_0\Gamma_7
		\nonumber\\
		&\quad
		- \lambda_x (\partial_x\phi)\,\tau_z\Gamma_8
		- \lambda_y (\partial_y\phi)\,\tau_0\Gamma_9 ,
		\label{Eq_Heff}
	\end{align}
	where $\Gamma_7=\sigma_z s_z$, $\Gamma_8=\sigma_x s_0$, and $\Gamma_9=\sigma_y s_z$ extend the matrices of Sec.~\ref{Sec:II}, with the renormalized kinetic and pairing terms
	\begin{align}
		\xi_{k,\mathrm{eff}} &= \xi_k + \frac{t}{4}\,
		|\boldsymbol{\nabla}\phi|^2 ,
		\\
		\Delta_{d,\mathrm{eff}} &= \left\{\Delta_d + \frac{\bar{\Delta}_0}{4}
		\left[(\partial_x\phi)^2 - (\partial_y\phi)^2\right]\right\}\,\tau_y\Gamma_3 
		\nonumber\\
		&\quad+ \bar{\Delta}_0
		\left[(\partial_x\phi)k_x - (\partial_y\phi)k_y\right]\tau_x\Gamma_2 .
		\label{Eq_Ddeff}
	\end{align}
		Equation~(\ref{Eq_Heff}) is the analogue of the effective model of 
	Ref.~\cite{Minakshi2025}, with two essential differences. First, 
	the constant pitch vector $\mathbf{g}$ of the spiral texture is replaced 
	by the local gradient $\boldsymbol{\nabla}\phi(\mathbf{r})$ of the 
	vortex--antivortex configuration, so that all texture-induced terms 
	acquire a spatial dependence. Second, the pairing amplitude 
	$\bar{\Delta}_0$ is not a free parameter but is determined 
	self-consistently from the gap equation~(\ref{Eq_GapEq}). The term 
	proportional to $\tau_0\Gamma_7$ describes a synthetic SOC 
	generated by the texture gradient, while in Eq.~(\ref{Eq_Ddeff}) the intrinsic $d$-wave pairing acquires an emergent $p$-wave contribution of the form $(\partial_x\phi)k_x-(\partial_y\phi)k_y$, which vanishes for a uniform texture. For the meron--antimeron pair configuration, $|\boldsymbol{\nabla}\phi|$ is largest near the texture cores and is of order $1/\min(N_x,N_y)$ along the system boundary, so the texture-induced terms are small corrections to the edge physics derived below.
	\subsection{Effective bulk pairing}
	\label{App_D2}
	
	To identify the pairing channels responsible for stabilizing the SOTSC 
	phase, we rewrite Eq.~(\ref{Eq_Heff}) such that the roles of kinetic energy and pairing are exchanged. A duality transformation turns the original $d$-wave pairing into a dual kinetic energy, while the mass and the spin--orbit terms of the QSHI become an effective pairing. We first change to a Nambu basis in which the two spin--orbit terms share the same 
	particle-hole structure, namely
	\begin{equation}
		\tilde{\Psi}_{\mathbf r_i} = W\,\Psi_{\mathbf r_i}
		= \left(c_{a\uparrow}, c_{a\downarrow}, c_{b\uparrow}, c_{b\downarrow}, 
		-c^{\dagger}_{a\downarrow}, c^{\dagger}_{a\uparrow}, 
		-c^{\dagger}_{b\downarrow}, c^{\dagger}_{b\uparrow}\right)^{T}_{\mathbf r_i},
		\label{Eq_basis_tilde}
	\end{equation}
	where $\Psi_{\mathbf r_i}$ is the Nambu spinor of Sec.~\ref{Sec:II} and $W = \mathrm{diag}\!\left(\mathbf{1}_4,\, \sigma_0\otimes(-is_y)\right)$ is a momentum-independent unitary transformation acting on the hole sector. Since $W$ is unitary, the two bases give the same spectrum. 
	In this basis the effective Hamiltonian~(\ref{Eq_Heff}) becomes $\tilde{H}_{\mathrm{eff}} = W H_{\mathrm{eff}} W^{\dagger}$, i.e.
	\begin{align}
		\tilde{H}_{\mathrm{eff}} 
		&= \xi_{k,\mathrm{eff}}\,\tau_z\Gamma_1 + \tilde{\Delta}_{d}\,\tau_x\Gamma_0
		+ J\,\tau_0\Gamma_2 
		\nonumber\\
		&\quad
		+ 2\lambda_x k_x\,\tau_z\Gamma_5 + 2\lambda_y k_y\,\tau_z\Gamma_6
		- t\left(\boldsymbol{\nabla}\phi\cdot\mathbf{k}\right)\tau_z\Gamma_7
		\nonumber\\
		&\quad
		- \lambda_x (\partial_x\phi)\,\tau_z\Gamma_8
		- \lambda_y (\partial_y\phi)\,\tau_z\Gamma_9
		\nonumber\\
		&\quad
		+ \bar{\Delta}_0\left[(\partial_y\phi)k_y - (\partial_x\phi)k_x\right]\tau_x\Gamma_4 ,
		\label{Eq_Heff_tilde}
	\end{align}
	where $\Gamma_0=\sigma_0 s_0$, $\Gamma_4=\sigma_0 s_z$, and 
	\begin{equation}
	\tilde{\Delta}_d=\Delta_d+\tfrac{\bar{\Delta}_0}{4}\big[(\partial_x\phi)^2-(\partial_y\phi)^2\big].
	\end{equation}
	 The pairing terms of Eq.~(\ref{Eq_Ddeff}) now appear as the particle-hole off-diagonal $\tau_x$ terms, and the exchange term becomes $\tau_0\Gamma_2$.
	
	Retaining terms up to linear order in momentum and performing the 
	duality transformation $U_d = \bar{\tau}\otimes\sigma_0\otimes s_0$ with
	\begin{equation}
		\bar{\tau} = \frac{1}{\sqrt{2}}
		\begin{pmatrix} 1 & -1 \\ 1 & 1 \end{pmatrix},
	\end{equation}
	the dual Hamiltonian $H_D = U_d^{\dagger}\tilde{H}_{\mathrm{eff}}U_d$ 
	takes the block form
	\begin{equation}
		H_D = 
		\begin{pmatrix}
			\Delta_D + J\sigma_0 s_x & \bar{\Delta}_D \\
			\bar{\Delta}_D & -\Delta_D + J\sigma_0 s_x
		\end{pmatrix},
		\label{Eq_HD}
	\end{equation}
	where 
	$\Delta_D = \bar{\Delta}_0
	\left[(\partial_y\phi)k_y - (\partial_x\phi)k_x\right]\sigma_0 s_z$ 
	plays the role of the dual kinetic energy and the off-diagonal block 
	$\bar{\Delta}_D$ represents the effective pairing. The latter decomposes 
	into two channels, $\bar{\Delta}_D = \bar{\Delta}_{s}^{\mathrm{eff}} + \bar{\Delta}_{p}^{\mathrm{eff}}$, 
	with
	\begin{align}
		\bar{\Delta}_{s}^{\mathrm{eff}} &= 
		-\left[(\epsilon_0 - 4t) + \frac{t}{4}|\boldsymbol{\nabla}\phi|^2
		\right]\sigma_z s_0 
		\nonumber\\
		&\quad
		+ \lambda\left[(\partial_x\phi)\,\sigma_x s_0 
		+ (\partial_y\phi)\,\sigma_y s_z\right],
		\label{Eq_Ds}
		\\
		\bar{\Delta}_{p}^{\mathrm{eff}} &= 
		-2\lambda\left(k_x\,\sigma_x s_z + k_y\,\sigma_y s_0\right) 
		+ t\left(\boldsymbol{\nabla}\phi\cdot\mathbf{k}\right)
		\sigma_z s_z ,
		\label{Eq_Dp}
	\end{align}
	where we have set $\lambda_x = \lambda_y \equiv \lambda$. The momentum-independent part $\bar{\Delta}_{s}^{\mathrm{eff}}$ is an effective $s$-wave pairing, while $\bar{\Delta}_{p}^{\mathrm{eff}}$ is a $p$-wave pairing: the $p_x+ip_y$-type term comes from the intrinsic SOC, and the $p_x+p_y$-type term from the finite texture gradient $\boldsymbol{\nabla}\phi$. Together, these channels stabilize the SOTSC phase hosting MCMs.
	\subsection{Low-energy edge theory}
	\label{App_D3}
	
	We now derive the effective edge Hamiltonian governing the boundary 
	physics and the tunability of the MCMs. We label the four edges of the 
	2D domain (Fig.~\ref{fig:Fig1}) as I ($y=0$), II ($x=N_x$), III ($y=N_y$), and IV ($x=0$). We neglect the texture-gradient terms ($\boldsymbol{\nabla}\phi\to 0$) and construct the edge theory from the uniform part of $\tilde{H}_{\mathrm{eff}}$ in Eq.~(\ref{Eq_Heff_tilde}).
	
	We start with edge I. We take OBC along $y$ and PBC along $x$, so that $k_x$ remains a good quantum number and $k_y\to -i\partial_y$. For $J,\bar{\Delta}_0 \ll |m'|, t, \lambda$, the exchange field and the pairing only perturb the QSHI. We therefore first solve for the helical edge states of the QSHI alone, and then project the exchange field and pairing onto them. For this, we partition $\tilde{H}_{\mathrm{eff}} = H_0 + H_p$~\cite{zhong2018,A_Ghosh_2021}, where the unperturbed part is the QSHI
	\begin{equation}
		H_0(-i\partial_y) = \left(m' - t\,\partial_y^2\right)
		\tau_z\Gamma_1 - 2i\lambda_y\partial_y\,\tau_z\Gamma_6 ,
		\label{Eq_H0_edge}
	\end{equation}
	where $m' = \epsilon_0 - 4t$ is the QSHI mass at $\mathbf{k}=0$. For $m'<0$ the mass $m'+tk^2$ changes sign between $k=0$ and large $k$, i.e., the bands are inverted~\cite{FuKane_2007}. This is the condition for $H_0$ to host edge states, as shown below. Further, $H_p$ collects the exchange field, the SOC along the edge, and the $d$-wave pairing:
	\begin{equation}
		H_p = J\,\tau_0\Gamma_2 + 2\lambda_x k_x\,\tau_z\Gamma_5 
		+ \bar{\Delta}_0\left(k_x^2 + \partial_y^2\right)
		\tau_x\Gamma_0 ,
		\label{Eq_Hp_edge}
	\end{equation}
	Solving $H_0\,\psi = 0$ with the boundary condition 
	$\psi(0) = \psi(\infty) = 0$ yields four zero-energy edge states
	\begin{equation}
		\psi_{\alpha}(y) = \mathcal{C}\, e^{-\kappa_1 y}
		\sin(\kappa_2 y)\, e^{ik_x x}\,\chi_{\alpha} ,
		\label{Eq_edge_wf}
	\end{equation}
	with $\kappa_1 = \lambda_y/t$, 
	$\kappa_2 = \sqrt{|m'|/t - \kappa_1^2}$ (both roots $\kappa_1\pm i\kappa_2$ decay into the bulk only for $m'<0$; for $m'>0$ one root grows, and $\psi(0)=0$ cannot be satisfied, so there are no edge states), normalization 
	$\mathcal{C} = 2\sqrt{\kappa_1(\kappa_1^2+\kappa_2^2)/\kappa_2^2}$, and 
	$\chi_{\alpha}$ the four spinors
	\begin{equation}
		\chi_{\alpha} = \eta_{\tau}\otimes\frac{1}{\sqrt{2}}\begin{pmatrix}1\\1\end{pmatrix}\!\otimes\,\eta_{s},
		\qquad \eta_{\tau},\eta_{s}\in\left\{\begin{pmatrix}1\\0\end{pmatrix},\begin{pmatrix}0\\1\end{pmatrix}\right\},
		\label{Eq_edge_spinor}
	\end{equation}
	i.e., the $\sigma_x=+1$ eigenstate in the orbital space, times any particle-hole and spin state. Projecting $H_p$ onto this edge subspace, 
	$\left(H^{j}_{\mathrm{edge}}\right)_{\alpha\beta} = 
	\int_0^{\infty} \psi^{\dagger}_{\alpha}\, H_p\, \psi_{\beta}\, dy$, 
	and repeating the construction for the remaining edges (with boundary 
	conditions $\psi(0)=\psi(-\infty)=0$ for edges III and II), the edge 
	Hamiltonians take the compact form
	\begin{equation}
		H^{j}_{\mathrm{edge}} = A_j\,\tau_x s_0 + C_j\,\tau_z s_z 
		+ E_j\,\tau_0 s_x ,
		\qquad j = \mathrm{I,\dots,IV},
		\label{Eq_edge_general}
	\end{equation}
	with coefficients
	\begin{align}
		A_j &= \left\{\frac{\bar{\Delta}_0 m'}{t},\,
		-\frac{\bar{\Delta}_0 m'}{t},\,
		\frac{\bar{\Delta}_0 m'}{t},\,
		-\frac{\bar{\Delta}_0 m'}{t}\right\},
		\nonumber\\
		C_j &= \left\{2\lambda k,\, 2\lambda k,\, 
		-2\lambda k,\, -2\lambda k\right\},
		\qquad
		E_j = \left\{J,\, 0,\, J,\, 0\right\},
		\label{Eq_edge_coeffs}
	\end{align}
	where $k$ denotes the momentum along the respective edge and we set 
	$\lambda_x = \lambda_y \equiv \lambda$. Two features of 
	Eq.~(\ref{Eq_edge_coeffs}) determine the corner physics. First, the 
	superconducting mass $A_j$ alternates sign between adjacent edges, 
	reflecting the $d_{x^2-y^2}$ form factor, which changes sign between 
	boundaries normal to $\hat{x}$ and $\hat{y}$. Second, the uniform 
	exchange (pointing along $\hat{x}$ in the rotated frame) projects onto 
	the edge states only for edges I and III.
	
	Setting $k=0$, the edge spectra read
	\begin{equation}
		E_{\mathrm{II,IV}} = \pm\frac{\bar{\Delta}_0 |m'|}{t},
		\qquad
		E_{\mathrm{I,III}} = \pm J \pm \frac{\bar{\Delta}_0 |m'|}{t} .
		\label{Eq_edge_spectra}
	\end{equation}
	Edges II and IV remain gapped as long as $\bar{\Delta}_0\neq0$, whereas the gap on edges I 
	and III closes at the first critical coupling $J_{c1}$, when
	\begin{equation}
		J_{c1} = \frac{\bar{\Delta}_0\, |m'|}{t} .
		\label{Eq_Jc_condition}
	\end{equation}
	Crucially, in our self-consistent framework $\bar{\Delta}_0$ is itself 
	a function of the system parameters, including $J$. The critical 
	exchange coupling separating the eight- and four-MCM phases is 
	therefore determined by the \emph{implicit} equation
	\begin{equation}
		J_{c1} = \frac{\bar{\Delta}_0\!\left(J_{c1}, \lambda, 
			\epsilon_0\right) |m'|}{t} ,
		\label{Eq_Jc_implicit}
	\end{equation}
	in contrast to the fixed-pairing case, where the right-hand side is 
	constant. Since $\bar{\Delta}_0$ does not increase with $J$ 
	[Fig.~\ref{fig:Fig.2}], the self-consistency systematically lowers 
	$J_{c1}$ relative to a fixed-amplitude estimate evaluated at the 
	$J = 0$ pairing strength. Beyond the transition, the sign change of 
	the total edge mass between adjacent edges binds Majorana states at 
	the corners via the Jackiw--Rebbi mechanism~\cite{Jackiw_Rebbi_1976}, as detailed in the next 
	subsection.
	%======================================================
	\subsection{Majorana corner mode solutions}
	The edge Hamiltonians~(\ref{Eq_edge_general}) have zero-energy corner solutions, which reveal the mechanism behind the tunable number of MCMs. We focus on the corner formed by edges I and II; the remaining corners follow analogously. Throughout this subsection we write $|m'| = 4t - \epsilon_0$ and $\Delta_m \equiv \bar{\Delta}_0|m'|/t$, and we use the edge-projected basis of Sec.~\ref{App_D3}, in which the spinors carry the particle-hole ($\tau$) and spin ($s$) indices, ordered as $(\tau_z,s_z)=(+,\uparrow),(+,\downarrow),(-,\uparrow),(-,\downarrow)$.

	Let $\ell$ be the coordinate along the boundary through the corner, with $\ell<0$ on edge I and $\ell>0$ on edge II. From Eqs.~(\ref{Eq_edge_general}) and~(\ref{Eq_edge_coeffs}), the edge Hamiltonian along this path is
	\begin{equation}
		H(\ell) = 2\lambda\,\tau_z s_z\,(-i\partial_\ell) + A(\ell)\,\tau_x s_0 + J\,\Theta(-\ell)\,\tau_0 s_x,
		\label{Eq_H_corner}
	\end{equation}
	with $A=-\Delta_m$ for $\ell<0$ and $A=+\Delta_m$ for $\ell>0$. The mass $A$ changes sign at the corner, while the exchange term is present only on edge I. The two mass terms commute with each other and both anticommute with $\tau_z s_z$. Hence Eq.~(\ref{Eq_H_corner}) separates into two sectors, $\tau_x s_x=\pm1$, each a one-dimensional Dirac Hamiltonian with mass
	\begin{equation}
		m_\pm(\ell) = A(\ell) \pm J\,\Theta(-\ell).
		\label{Eq_mass_sectors}
	\end{equation}
	A zero-energy solution of the Jackiw--Rebbi type~\cite{Jackiw_Rebbi_1976} is
	\begin{equation}
		\psi_\pm(\ell) = \exp\!\left[-\frac{1}{2\lambda}\int_0^\ell m_\pm(\ell')\,d\ell'\right]\phi_\pm,
		\label{Eq_JR}
	\end{equation}
	where $\phi_\pm$ is the eigenspinor with $\tau_y s_z=-1$ in the sector $\tau_x s_x=\pm1$,
	\begin{equation}
		\phi_+ = (1,-i,-i,1)^{T},
		\qquad
		\phi_- = (1,i,-i,-1)^{T}.
		\label{Eq_corner_spinors}
	\end{equation}
	The solution is a corner mode if it decays on both sides of the corner, i.e., if $m_\pm$ is positive on edge II and negative on edge I.

	On edge II, $m_\pm=\Delta_m>0$ in both sectors, so both solutions decay as $e^{-\Delta_m\ell/2\lambda}$, i.e., with the decay constant
	\begin{equation}
		\kappa_{\mathrm{II}} = \frac{\bar{\Delta}_0|m'|}{2\lambda t}.
		\label{Eq_kappa_edgeII}
	\end{equation}
	On edge I, $m_\pm=-\Delta_m\pm J$, which gives the decay constants
	\begin{equation}
		\kappa_{\mathrm{I}}^{\pm} = \frac{\bar{\Delta}_0|m'| \mp Jt}{2\lambda t}.
		\label{Eq_kappa_edgeI}
	\end{equation}
	The solution $\psi_-$ always decays on edge I, since $\kappa_{\mathrm{I}}^{-}>0$. The solution $\psi_+$ decays on edge I only if $\kappa_{\mathrm{I}}^{+}>0$, i.e., if $J<\bar{\Delta}_0|m'|/t$. This gives two regimes.

	\emph{Case I: $J > \bar{\Delta}_0|m'|/t$ (four-MCM phase).} Only $\psi_-$ is normalizable, and each corner hosts a single Majorana mode,
	\begin{align}
		\psi_C &\propto 
		\exp\!\left[-\frac{\bar{\Delta}_0|m'|}{2\lambda t}\,y\right]
		\phi_-
		\quad \text{(along edge II)},
		\nonumber\\
		\psi_C &\propto 
		\exp\!\left[-\frac{Jt + \bar{\Delta}_0|m'|}{2\lambda t}\,|x-N_x|\right]
		\phi_-
		\quad \text{(along edge I)},
		\label{Eq_MCM_caseI}
	\end{align}
	with anisotropic localization lengths $\xi_y = 2\lambda t/(\bar{\Delta}_0|m'|)$ and $\xi_x = 2\lambda t/(Jt + \bar{\Delta}_0|m'|)$. With one mode per corner, the system hosts four MCMs in total.

	\emph{Case II: $J < \bar{\Delta}_0|m'|/t$ (eight-MCM phase).} Both $\psi_+$ and $\psi_-$ are normalizable. Each corner then hosts two Majorana modes, $\psi_C=\psi_+$ and $\psi_C=\psi_-$, where along edge I
	\begin{align}
		\psi_{\pm} &\propto 
		\exp\!\left[-\frac{\bar{\Delta}_0|m'| \mp Jt}{2\lambda t}\,|x-N_x|\right]
		\phi_{\pm},
		\label{Eq_MCM_caseII}
	\end{align}
	and along edge II both decay as in Eq.~(\ref{Eq_MCM_caseI}). The two modes have different localization lengths along edge I, controlled by $\bar{\Delta}_0|m'| \mp Jt$. With two modes per corner, the system hosts eight MCMs in total.

	The transition between the two cases occurs precisely at the edge 
	gap-closing condition~(\ref{Eq_Jc_condition}), so the analytic corner 
	counting reproduces the eight-to-four MCM transition observed 
	numerically in Fig.~\ref{fig:Fig3}. As $J \to J_{c1}^{-}$, the second 
	localization length in Eq.~(\ref{Eq_MCM_caseII}) diverges, 
	$\xi \sim 2\lambda t/(\bar{\Delta}_0|m'| - Jt) \to \infty$: one member 
	of each corner pair delocalizes along the edge and gaps out, leaving a 
	single MCM per corner. Within the self-consistent framework, 
	$\bar{\Delta}_0 = \bar{\Delta}_0(J)$ in all expressions above, so both 
	the localization lengths and the transition point inherit the 
	$J$ dependence of the pairing amplitude. This analysis also accounts 
	for the geometry dependence found in Appendix~\ref{App_B}: corner 
	truncations that remove an intersection of mass domains remove the 
	associated Majorana pair, reducing the count from eight to four and 
	from four to two in the triangular geometries.

	\bibliography{references.bib}
	%======================================================
\end{document}